\documentclass[fleqn,usenatbib]{mnras}

\usepackage{newtxtext,newtxmath}

\usepackage[T1]{fontenc}

\DeclareRobustCommand{\VAN}[3]{#2}
\let\VANthebibliography\thebibliography
\def\thebibliography{\DeclareRobustCommand{\VAN}[3]{##3}\VANthebibliography}

\usepackage{graphicx}	
\usepackage{amsmath}	
\usepackage{pifont}
\usepackage{mathtools}
\usepackage{placeins}
\usepackage{float}
\usepackage{bm}
    
\usepackage{nccmath}

\usepackage{cleveref}
\crefname{figure}{Fig.}{Figs.}
\crefname{table}{Table}{Tables}
\usepackage{amsmath}
\usepackage{units}
\usepackage[percent]{overpic}
\usepackage{subfigure}
\usepackage{longtable}
\usepackage{threeparttable}
\usepackage{appendix}
\usepackage[figuresright]{rotating}
\usepackage{lscape}
\usepackage{ulem}
\defcitealias{2022ApJ...938...42H}{H22}

\title[]{The role of tidal interactions in the formation of slowly rotating early-type stars in young star clusters}

\author[C. He et al.]{Chenyu He,$^{1,2}$\thanks{E-mail:
    hechy9@mail2.sysu.edu.cn} Chengyuan Li,$^{1,2}$ Weijia Sun,$^{3}$
  Richard de Grijs,$^{4,5}$ Lu Li,$^{6}$ Jing Zhong,$^{6}$ Songmei
  Qin,$^{6,7}$ \newauthor Li Chen,$^{6,7}$ Li Wang,$^{1,2}$ Baitian
  Tang,$^{1,2}$ Zhengyi Shao$^{6,8}$ and Cheng Xu $^{1,2}$
\\
\\
$^{1}$School of Physics and Astronomy, Sun Yat-sen University, Zhuhai, 519082, China\\
$^{2}$CSST Science Center for the Guangdong--Hong Kong--Macau Greater Bay Area, Zhuhai, 519082, China\\
$^{3}$Leibniz-Institut für Astrophysik Potsdam (AIP), Germany\\
$^{4}$School of Mathematical and Physical Sciences, Macquarie University, Sydney, NSW 2109, Australia\\
$^{5}$Astrophysics and Space Technologies Research Centre, Macquarie University, Sydney, NSW 2109, Australia\\
$^{6}$Key Laboratory for Research in Galaxies and Cosmology, Shanghai Astronomical Observatory, Chinese Academy of Sciences, 80 Nandan Road,\\   Shanghai, 200030, China\\
$^{7}$School of Astronomy and Space Science, University of Chinese Academy of Sciences, 19A Yuquan Road, Beijing, 100049, China\\
$^{8}$Key Laboratory for Astrophysics, Shanghai, 200234, People's Republic of China\\
}

\date{Accepted XXX. Received YYY; in original form ZZZ}

\pubyear{2023}

\begin{document}
\label{firstpage}
\pagerange{\pageref{firstpage}--\pageref{lastpage}}
\maketitle

\begin{abstract}
The split main sequences found in the colour--magnitude diagrams of star clusters younger than $\sim \unit[600]{Myr}$ are suggested to be caused by the dichotomy of stellar rotation rates of upper main-sequence stars. Tidal interactions have been suggested as a possible explanation of the dichotomy of the stellar rotation rates. This hypothesis proposes that the slow rotation rates of stars along the split main sequences are caused by tidal interactions in binaries. To test this scenario, we measured the variations in the radial velocities of slowly rotating stars along the split main sequence of the young Galactic cluster NGC 2422 ($\sim \unit[90]{Myr}$) using spectra obtained at multiple epochs with the Canada--France--Hawai'i Telescope. Our results show that most slowly rotating stars are not radial-velocity variables. Using the theory of dynamical tides, we find that the binary separations necessary to fully or partially synchronise our spectroscopic targets, on time-scales shorter than the cluster age, predict much larger radial velocity variations across multiple-epoch observations, or a much larger radial velocity dispersion at a single epoch, than the observed values. This indicates that tidal interactions are not the dominant mechanism to form slowly rotating stars along the split main sequences. As the observations of the rotation velocity distribution among B- and A-type stars in binaries of larger separations hint at a much stronger effect of braking with age, we discuss the consequences of relaxing the constraints of the dynamical tides theory.
\end{abstract}


\section{Introduction}

Star clusters younger than $\unit[2]{Gyr}$ in the Milky Way and the
Magellanic Clouds (MCs) are widely found to display extended
main-sequence turn-offs \citep[eMSTOs; e.g.,][]{2008ApJ...681L..17M,
  2009A&A...497..755M, 2009AJ....137.4988G,2011ApJ...737....4G,
  2018MNRAS.477.2640M,2018ApJ...869..139C,
  2019ApJ...876...65L}. Clusters younger than $\unit[600]{Myr}$
exhibit not only eMSTOs, but also a split pattern in their upper
main-sequence (MS) regions \citep[e.g.,][]{2016MNRAS.458.4368M,
  2018MNRAS.477.2640M, 2017MNRAS.467.3628C, 2017ApJ...844..119L,
  2019ApJ...883..182S}. Since they cannot be fitted by isochrones with a single age and a single metallicity, the notion that star clusters are 'simple stellar populations', i.e., that stars in a
cluster have similar ages and chemical content, is
challenged. However, eMSTOs and split MSs cannot be reproduced by
chemical differences in helium or metallicity
\citep{2016MNRAS.458.4368M}. Observation results also exclude an
extended star-formation history as long as that implied by the widths
of the eMSTOs; the latter is not consistent with the absence of gas
\citep{2014MNRAS.443.3594B} and the narrow morphologies of the
subgiant branches and/or tight red clumps
\citep{2014Natur.516..367L,2014ApJ...784..157L,2016MNRAS.461.3212L,
  2015MNRAS.448.1863B} in young massive MC clusters. Therefore, the
suggested existence of multiple stellar populations with different
ages and/or chemical variations in these clusters raises suspicions.

Many studies have found strong correlations between the eMSTOs and
split MSs with differences in stellar rotation rates in those regions
\citep[e.g.,][]{2009MNRAS.398L..11B,
  2018MNRAS.477.2640M,2018ApJ...869..139C,2019ApJ...876..113S,
  2019ApJ...883..182S}. The eMSTOs and split MSs are thus suggested to
have been caused by different stellar rotation rates in otherwise
`simple' stellar populations
\citep[e.g.,][]{2009MNRAS.398L..11B,2018ApJ...869..139C}. The eMSTOs
and split MSs in many young MC clusters are located at loci defined by
masses $M>\unit[1.6]{M_{\odot}}$
\citep{2018MNRAS.477.2640M}. \cite{2022ApJ...925..159Y} determined a
lower limit to the critical mass for the appearance of the eMSTO of
$\sim \unit[1.54]{M_{\odot}}$ in the Galactic cluster NGC 6819 ($\sim
\unit[2.5]{Gyr}$). These stellar masses are close to the onset mass
predicted theoretically, where magnetic braking of stellar rotation becomes significant \citep{1967ApJ...150..551K}. Therefore, this
indicates a possible correlation between a spread in stellar rotation
rates and the appearance of eMSTOs and/or split MSs. For eMSTOs,
spectroscopic studies have shown that stars on the red side have
larger average projected rotational velocities, $v\sin i$, than those
on the blue side
\citep{2017ApJ...846L...1D,2020MNRAS.492.2177K,2019ApJ...876..113S},
whereas simple stellar populations with single ages and different
rotation rates seem unable to reproduce the entire (e)MSTO stellar
distributions in some clusters
\citep[][]{2017MNRAS.465.4363M,2017MNRAS.467.3628C,2017ApJ...846...22G,2019ApJ...876...65L}. For split MSs, the blue and red MSs (bMS and rMS,
respectively) are well-fitted by isochrones populated by coeval slowly
and rapidly rotating populations, respectively
\citep{2015MNRAS.453.2637D, 2016MNRAS.458.4368M,
  2017NatAs...1E.186D}. Spectroscopic absorption-line profiles of
split-MS stars reveal a clear difference between the $v\sin{i}$ of the
bMS and rMS stars in several Large MC (LMC) and Galactic clusters
\citep{2018AJ....156..116M,
  2018ApJ...863L..33M,2019ApJ...883..182S,2023MNRAS.518.1505K}. Intriguingly,
the detected $v\sin{i}$ distributions of the split MSs seem to be
bimodal \citep{2018AJ....156..116M,
  2018ApJ...863L..33M,2019ApJ...883..182S,2023MNRAS.518.1505K}, which
is thought to be related to the dichotomy of their equatorial rotation
rates \citep{2019ApJ...883..182S}. This bimodality for
the equatorial rotation rates has also been found in massive ($M>\unit[2.5]{M_\odot}$) field stars \citep{2012A&A...537A.120Z,2021ApJ...921..145S}, whose appearance is found to be related to stellar metallicity \citep{2021ApJ...921..145S}. \cite{2013A&A...550A.109D} also found a bimodal equatorial rotation rate distribution in single
early B-type filed stars.

The reason as to why the distribution of stellar rotation rates is
dichotomous remains unresolved. \citet{2020MNRAS.495.1978B} suggested
that massive stars might have a bimodal rotation-period distribution
at the pre-main-sequence (PMS) stage. This bimodality is retained onto
the MS, resulting in a spread of stellar rotation rates among MS
stars. They suggested that star--disc interactions (disc locking)
might play a role to form such a double-peaked period distribution,
i.e., stars experiencing long (short) disc-braking time-scales during
the PMS form slow (rapid) rotators. \citet{2022NatAs...6..480W}
suggested that the slow rotators on the bMSs may be the product of
binary mergers. The rejuvenation and higher mass of the newly formed
core hydrogen content makes them appear bluer than single stars of the
same age and mass. Loss of angular momentum during thermal expansion
after coalescence and internal restructuring causes the slow rotation
of the merger product \citep{2019Natur.574..211S}. This model could
reproduce all properties observed at the eMSTOs and is supported by
the different stellar mass functions of the bMS stars compared to
those of the rMS stars as derived in some young MC clusters
\citep{2022NatAs...6..480W}.

\citet{2015MNRAS.453.2637D} suggested that tidal interactions in binaries may be responsible for the slow rotation of bMS stars. This mechanism has been suggested by \cite{2012A&A...537A.120Z} to possibly account for the slow rotation of massive field stars. \cite{2017NatAs...1E.186D} suggested a braking model. They proposed that all split MS stars are fast rotators at birth; the rotation rates of some stars are then braked, forming a bMS that is separated from the rapidly rotating rMS. They proposed that tidal interactions could be a possible mechanism to introduce such braking. Based on the observation of \cite{2004ApJ...616..562A}, \citet{2015MNRAS.453.2637D} suggested that the slowly rotating split MS stars could be components of binaries with periods of 4 to 500 days. They also recommended to examine the dynamical tides theory, which was suggested by \cite{1975A&A....41..329Z,1977A&A....57..383Z} to account for the tidal braking of the rotation of stars with radiative envelopes in close binaries.

Based on the dynamical tides model, the slowly rotating bMS stars should have close companions. However, \citet{2015MNRAS.453.2637D} noticed that an important fraction of binaries are circularised even if the semi-major axis is larger than predicted by the dynamical tides model. This indicates that other mechanisms may also be efficient at braking the rotation of massive stars in binaries. By exploring the binarity of stars using the variability in their radial velocities, \cite{2020MNRAS.492.2177K} and
\citet{2021MNRAS.508.2302K} detected similar binary fractions among
the slowly and rapidly rotating stars at the eMSTO of NGC 1846 ($\sim
\unit[1.5]{Gyr}$) and along the split MS of NGC 1850 ($\sim
\unit[100]{Myr}$), respectively. In \cite{Wang_2023}, we explored the role of tidal interaction in the formation of bMS stars in an NGC 1856 \citep[$\sim$ 300
  Myr;][]{2005ApJS..161..304M}-like mock cluster using $N$-body
simulations. We found that only high-mass-ratio binaries can be
tidally locked on time-scale on the order of the age of NGC 1856, where the dynamical tides model was applied in the binary evolution models \citep{2002MNRAS.329..897H}. However, they are located close to the equal-mass-ratio binary sequence, which
is much redder than the bMS. This indicates that tidal locking based on the dynamical tides model cannot account for the formation of the bMS stars. \cite{2021ApJ...912...27Y} revealed that the spatial distributions of the bMS stars in four MC clusters showed
a strong anti-correlation with those of high-mass-ratio binaries. This may not be consistent with the expectation from the scenario of tidal interactions for the formation of slowly rotating bMS stars. 

In this paper, we directly examine if most slowly rotating stars might hide a close
binary component along the split MS of NGC 2422 \citep[$\sim$90
  Myr;][hereafter
  \citetalias{2022ApJ...938...42H}]{2022ApJ...938...42H}, by measuring
the variations in their radial velocities (RVs). We aim to explore the tidal interaction scenario in the framework of the dynamical tides theory. We also discuss our results based on the observation of \cite{2004ApJ...616..562A}, who found that the rotation of the B0--F0 stars in binaries with periods of 4--500 days can also be significantly slowed down compared with single stars. We obtained multiple-epoch spectroscopic
observations with the Canada--France--Hawai'i Telescope (CFHT) for
these slow rotators. Combined with their spectra observed with the
same facility by \citetalias{2022ApJ...938...42H}, we measured their
RV differences at different epochs.

This paper is organised as follows. In Section \ref{sec:data_re} we
describe our data reduction method. In Section \ref{sec:result} we
show our RV measurement results and compare them with synthetic RV
variations and the dispersion expected from tidally locked binaries. In Section \ref{sec:discussion} we discuss our results, then
reach our conclusions in Section \ref{sec:conclusion}.

\section{Data Reduction}\label{sec:data_re}

In \citetalias{2022ApJ...938...42H}, we measured the projected
rotation rates of 47 split-MS stars in NGC 2422. These 47 stars were
selected based on their proper motions and are thus
RV-independent. This cluster is located in the Milky Way at a distance
of $\sim \unit[476]{pc}$. The $v\sin i$ values of the spectroscopic
targets span a large range, from $\unit[5]{km\,s^{-1}}$ to
$\unit[325]{km\,s^{-1}}$, where the average $v\sin i$ of the bMS stars
$\sim \unit[100]{km\,s^{-1}}$, with a standard deviation of $\sim
\unit[54]{km\,s^{-1}}$. To avoid contamination from rapid rotators, we
selected 21 stars with $v\sin i\leq \unit[104]{km\,s^{-1}}$ as our
sample of slowly rotating stars. We conducted time-domain
spectroscopic observations for these objects through CFHT programme
21BS005. Each star was observed one to three times with ESPaDOnS, with
a spectral resolution of $R\approx 68,000$ and a signal-to-noise ratio
(S/N) of $\sim 30$ at $\unit[4440]{\text{\AA}}$, thus covering the Mg
{\sc ii} $\unit[4481]{\text{\AA}}$ absorption line. The observations
were taken at two epochs. The first round of observations was obtained
during the period 26--27 November 2021, while the second ran from 27
to 28 December 2021. We also retrieved spectra from the CFHT programme
20BS002, which we used to measure the $v\sin i$ in
\citetalias{2022ApJ...938...42H}. The spectra of programme 20BS002
have the same resolution as those of programme 21BS005 but a different
S/N $\sim 45$ at $\unit[4440]{\text{\AA}}$. The spectra of CFHT
programme 20BS002 were observed during the period 21--27 November
2020. The reduced 1D spectra from both programmes were obtained after
processing by the CFHT pipeline, and the effects of the Earth's motion
have been removed.

In Table \ref{tab:fitted_lines} and Table \ref{tab:fitting_result} we include detailed information
pertaining to each observed star. The \textit{Gaia} IDs used in this
paper are from \textit{Gaia} Early Data Release 3
\citep[EDR3;][]{2016A&A...595A...1G,2021A&A...649A...1G}. The
corresponding epoch for each observation is included in the notes of Table \ref{tab:fitting_result}. Among our targets, 18 stars were observed at least twice. We
denote them as 2-Obs stars. Among the 2-Obs stars, a subsample of 12
stars were observed three times. We denote these as 3-Obs stars. Three
stars (\textit{Gaia} IDs: 3030028684933623808, 3030014013325413248 and
3030026138007337088) were observed only once under the two CFHT
programmes. They are denoted as 1-Obs stars. Their RVs will be
discussed when we compare the observed RV dispersion of the 21 targets
at a single epoch with those expected for synthetic tidally locked
binaries (see Section \ref{sec:result}).

\begin{figure*}
    \includegraphics[width=0.7\textwidth]{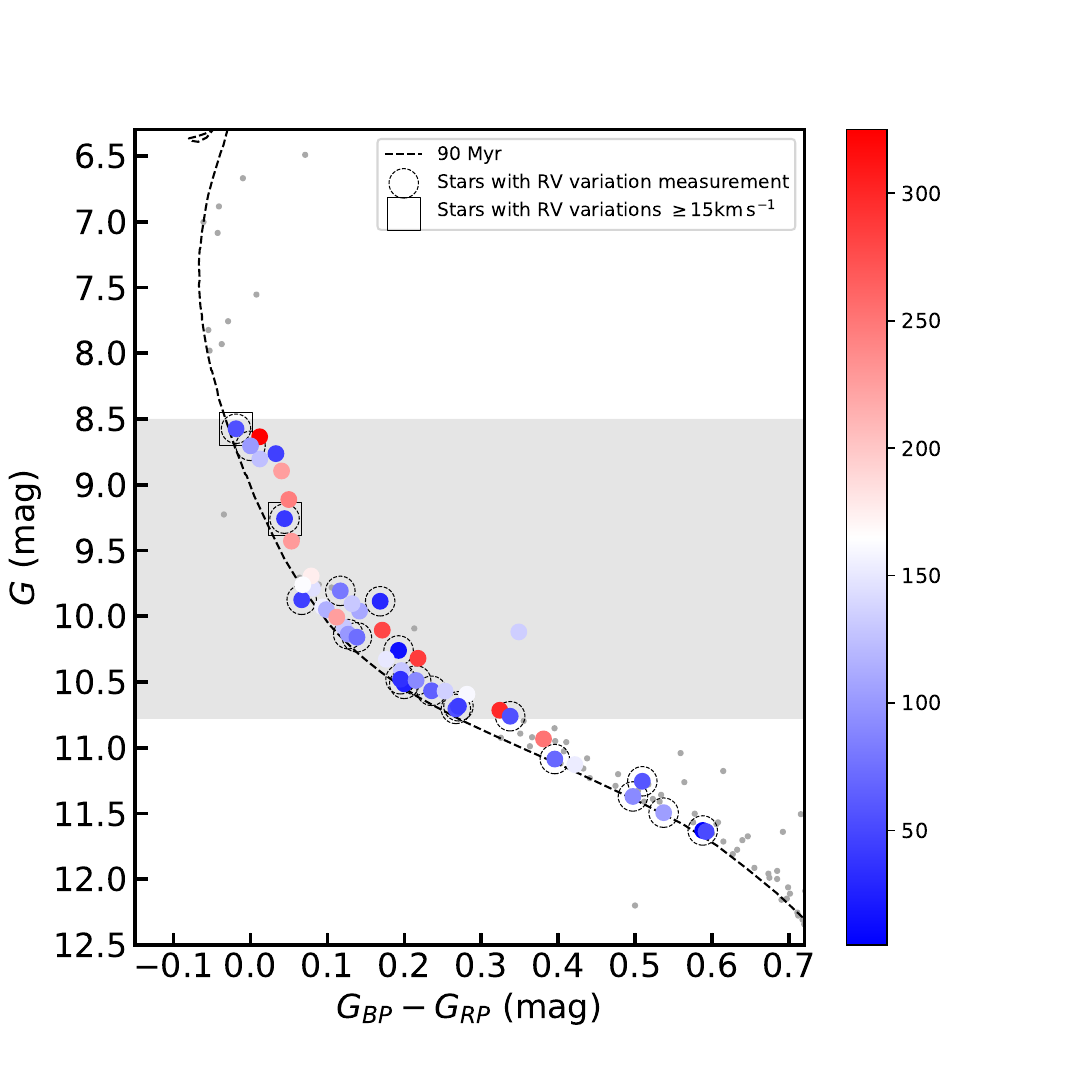}
	\caption{Loci of the 21 spectroscopic targets in the CMD of
          NGC 2422 with RV measurements used in this paper, encircled
          by dashed open circles. The coloured dots show the 47 stars
          explored spectroscopically by
          \citetalias{2022ApJ...938...42H}, colour-coded by the
          measured $v\sin i$ values from
          \citetalias{2022ApJ...938...42H}. The other cluster members
          (grey dots) and the best-fitting isochrone to the blue edge
          of the eMSTO (dashed line) were adopted from
          \citetalias{2022ApJ...938...42H}, where the theoretical
          isochrone is from the PARSEC models \citep[version
            1.2S,][]{2017ApJ...835...77M}. The photometric data are
          from \textit{Gaia} EDR3
          \citep{2016A&A...595A...1G,2021A&A...649A...1G}. The shading
          shows the region $\unit[8.5]{mag}<G<\unit[10.78]{mag}$ where
          most cluster members have been spectroscopically explored by \citetalias{2022ApJ...938...42H}. The RV variable candidates
          are shown as solid open squares.}\label{fig:RV_samples}
\end{figure*}

Figure~\ref{fig:RV_samples} shows the loci of the 21 spectroscopic
targets in the colour--magnitude diagram (CMD) of NGC 2422. In
Figure~\ref{fig:RV_samples}, the cluster member stars, the input of
the best-fitting isochrone to the blue edge of the cluster eMSTO and
the $v\sin i$ values were adopted from
\citetalias{2022ApJ...938...42H}. The photometric data are from
\textit{Gaia} EDR3, whereas the best-fitting isochrone was adopted
from the PARSEC model \citep[version
  1.2S;][]{2017ApJ...835...77M}. The 21 targets account for 91 per
cent (21/23) of the stars with $v\sin i\leq \unit[104]{km\,s^{-1}}$
detected by \citetalias{2022ApJ...938...42H}. The grey shading shows
the region spanning $\unit[8.5]{mag}<G<\unit[10.78]{mag}$, including
the loci where the split pattern is most evident. This corresponds to
an inferred mass range of $\unit[1.75]{M_\odot}<M<\unit[3.60]{M_\odot}$,
based on the best-fitting isochrone. Within this range, for 39 out of
47 member stars (81 per cent), spectra were obtained by
\citetalias{2022ApJ...938...42H}, and 17 stars have $v\sin i\leq
\unit[104]{km\,s^{-1}}$. The spectroscopic targets in this paper
comprise 94 per cent (16/17) of stars with $v\sin
i\leq\unit[104]{km\,s^{-1}}$ within the grey shaded area. Supposing
that all eight members not studied spectroscopically by
\citetalias{2022ApJ...938...42H} are slow rotators, we have a minimum
sample completeness of 64 per cent (16/25) for stars with $v\sin i\leq
\unit[104]{km\,s^{-1}}$ in the grey shaded area. Therefore, any bias
caused by the selection of stars for our RV measurements should not be
significant for our statistical analysis of the RV variability of slow
rotators within the grey shaded area. Fifteen 2-Obs and 11 3-Obs stars
are located within the grey shaded area. We denote them as 2-Obs-shade
and 3-Obs-shade stars, respectively.

\begin{figure*}
\begin{tabular}{cc}
    \includegraphics[width=0.4\textwidth]{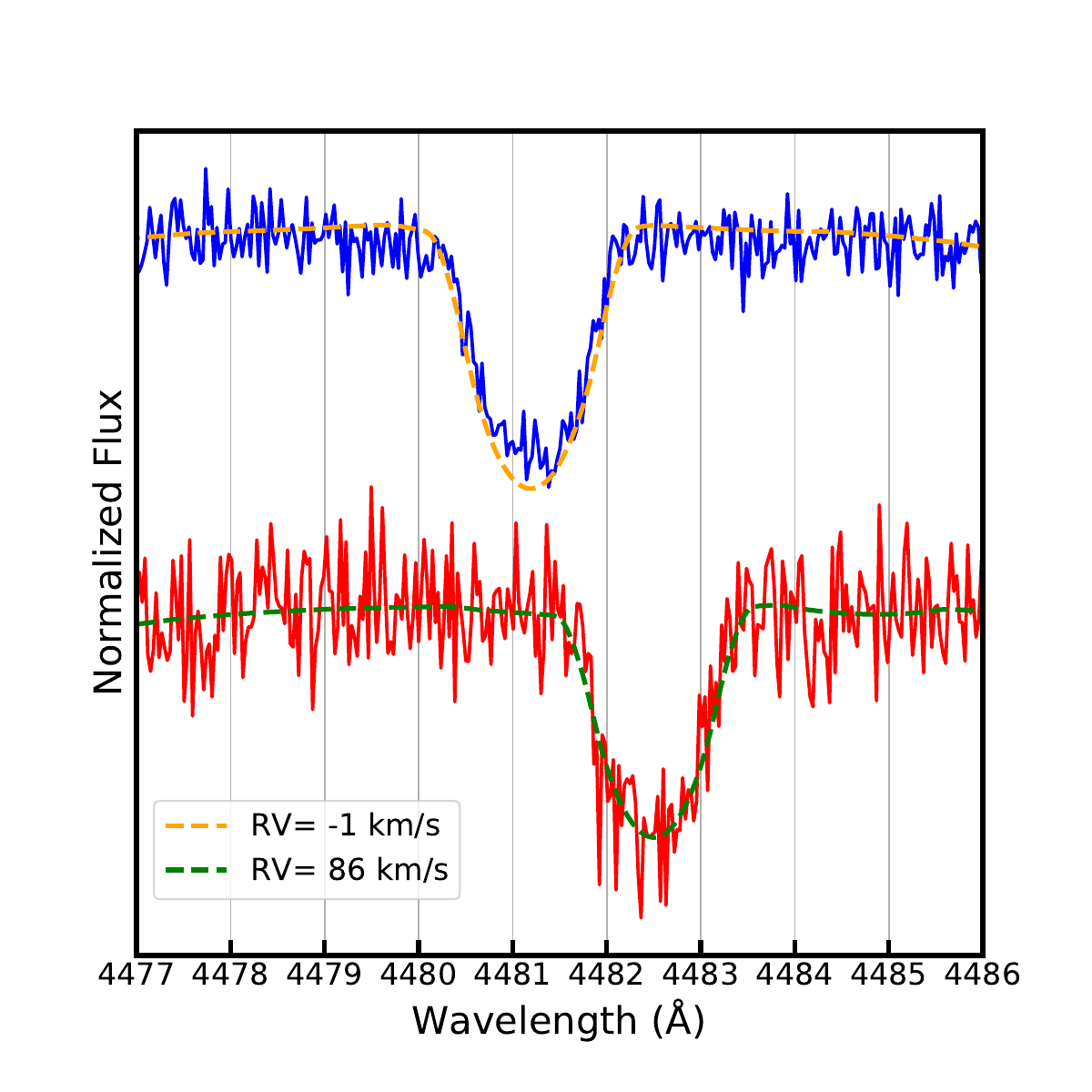}  & \includegraphics[width=0.4\textwidth]{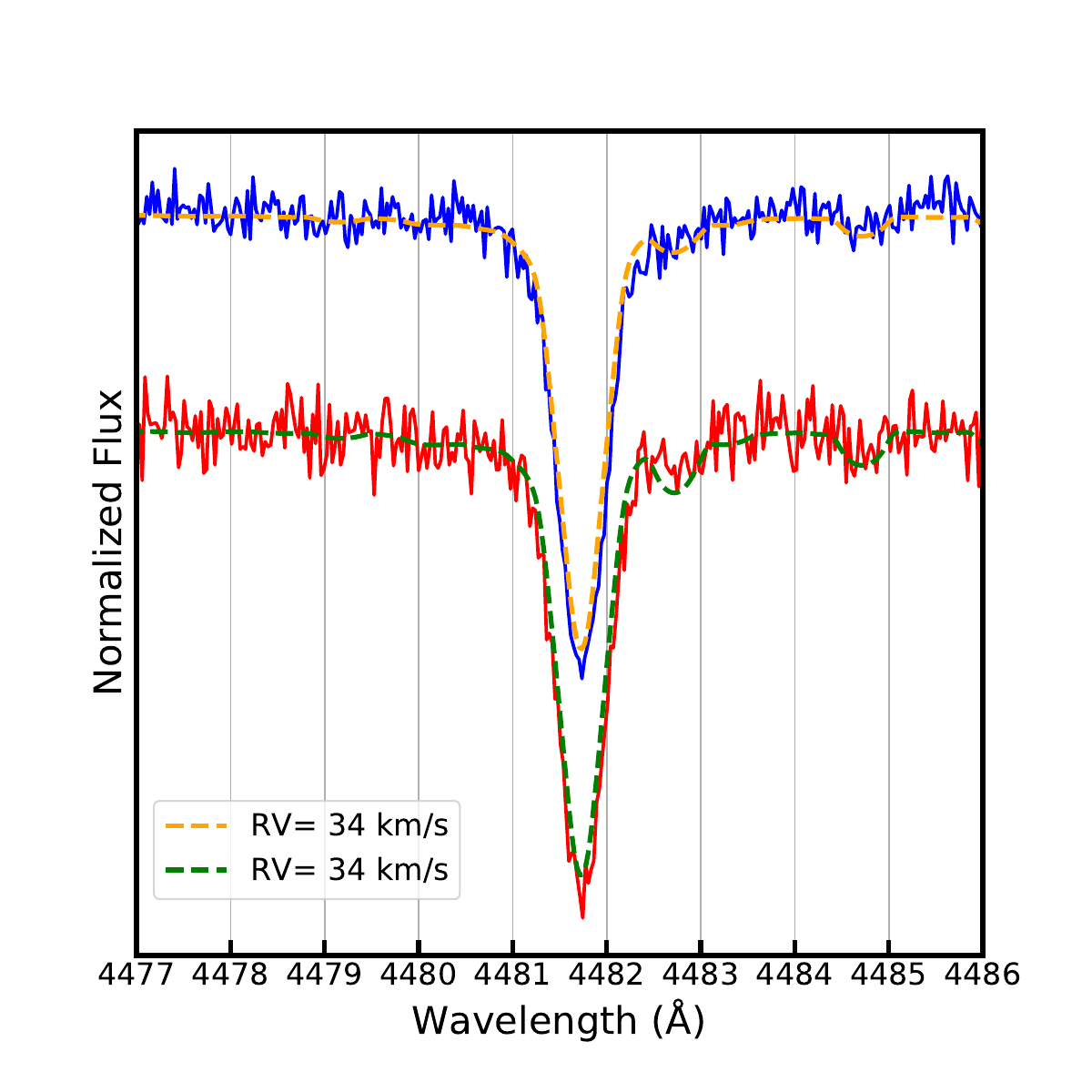}  \\
\end{tabular}
	\caption{Observed Mg {\sc ii} $\unit[4481.1]{\text{\AA}}$
          absorption profiles (solid lines) for \textit{Gaia} ID
          3030259479295155072 (left) and \textit{Gaia} ID
          3030298374519750912 (right), with their best-fitting models
          (dashed lines). In both panels, the blue and red spectra
          represent the observations obtained in November 2020 and
          December 2021, respectively. The input RVs of the
          best-fitting models are listed in the
          legend.}\label{fig:spectrum samples}
\end{figure*}

The RVs of the stars were measured by fitting the observed spectral
absorption-line profiles. We first derived synthetic stellar spectra
from the Pollux database \citep{2010A&A...516A..13P}. The synthetic
spectra were generated with the SYNSPEC tool
\citep{1992A&A...262..501H} based on the plane-parallel ATLAS12 model
atmospheres in local thermodynamic equilibrium
\citep{2005MSAIS...8..189K}, with a fixed microturbulent velocity of
$\unit[2]{km\,s^{-1}}$. The stellar effective temperatures
$T_\mathrm{eff}$ of the synthetic spectra range from $\unit[6,000]{K}$
to $\unit[15,000]{K}$, in steps of $\unit[100]{K}$, with the surface
gravities ranging from $\log g = 3.5$ dex to $\log g =5.0$ dex, in
steps of $\unit[0.1]{dex}$. The [Fe/H] abundances of the synthetic
spectra range from $\unit[-1.0]{dex}$ to $\unit[1.0]{dex}$, in steps
of $\unit[0.5]{dex}$; we fixed the [Fe/H] at $\unit[0.0]{dex}$, based
on the metallicity inferred from the best-fitting isochrone
\citepalias{2022ApJ...938...42H}. The synthetic spectra were convolved
with the effects of instrumental and rotational ($v\sin i$) broadening
using the PyAstronomy tool \citep{pya}. The $v\sin i$ input for each
star was set within its $v\sin i$ range from
\citetalias{2022ApJ...938...42H}, within $\pm \unit[20]{km\,s^{-1}}$
in steps of $\unit[5]{km\,s^{-1}}$. The wavelengths were shifted using
PyAstronomy based on RVs from $\unit[-50]{km\,s^{-1}}$ to
$\unit[100]{km\,s^{-1}}$, in steps of $\unit[1]{km\,s^{-1}}$. We next
used Astrolib's PySynphot \citep{2013ascl.soft03023S} to calculate the
corresponding flux of the model spectra for each wavelength of the
observed spectra. We generated a series of model spectra with
different $T_\mathrm{eff}$, $\log g$, $v\sin i$ and RVs and a fixed
[Fe/H] $=\unit[0]{dex}$ and fitted the observed profiles of at least
three absorption lines for each star. The absorption lines fitted for
our sample stars are summarised in Table \ref{tab:fitted_lines}. The
best-fitting model for each spectrum was determined using a
minimum-$\chi^{2}$ method. The uncertainties in the $v\sin i$ and RV
values were estimated by comparing the parameters of the best-fitting
models with a series of mock spectra with well-known parameters. In
Figure~\ref{fig:spectrum samples}, we show the multiple-epoch spectra
of two example stars along with their best-fitting models. One star
shows a large RV variation (left panel) and the other exhibits a small
RV variation (right panel).

\begin{table*}
	\centering
	\caption{List of Fitted Absorption Lines for Each Star\label{tab:fitted_lines}}
    \begin{threeparttable}
    \begin{tabular}{||c|c|l||}
  \hline\hline
  \multicolumn{1}{|c|}{\textit{Gaia} ID\tnote{a}} &
  \multicolumn{1}{c|}{$G$ (mag)} &
  \multicolumn{1}{|c|}{Lines fitted\tnote{b}} \\
\hline
  3030259479295155072 & 8.57 & Mg {\sc ii} (4481.1), Fe {\sc ii} (4549.5), Si {\sc ii} (6347.1, 6371.4) \\
  3030027447983067008 & 8.70 & He {\sc i} (4471.5), Mg {\sc ii} (4481.1), Si {\sc ii} (6347.1, 6371.4), O {\sc i} (7771--7776)\\
  3028387801268979584 & 9.26 & Mg {\sc ii} (4481.1), Fe {\sc ii} (4549.5, 5169.0), Si {\sc ii} (6371.4) \\
  3030231785345188608 & 9.81 & Mg {\sc ii} (4481.1), Fe {\sc ii} (4549.5, 5169.0), Mg {\sc i} (5167.3, 5172.7)\\
  3030030746517945600 & 9.88 & Mg {\sc ii} (4481.1), Fe {\sc ii} (4549.5, 5169.0), Mg {\sc i} (5167.3, 5172.7)\\
  3030250751920573696 & 9.89 & Mg {\sc ii} (4481.1), Fe {\sc ii} (4549.5, 5169.0), Mg {\sc i} (5167.3, 5172.7)\\
  3029232707231846784 & 10.14 & Mg {\sc ii} (4481.1), Fe {\sc ii} (4549.5, 5169.0), Mg {\sc i} (5167.3, 5172.7)\\
  3030038546178432256 & 10.16 & Mg {\sc ii} (4481.1), Fe {\sc ii} (4549.5, 5169.0), Mg {\sc i} (5167.3, 5172.7)\\
  3030298374519750912 & 10.26 & Mg {\sc ii} (4481.1), Fe {\sc ii} (4549.5, 5169.0), Ti {\sc ii} (4572.0), Mg {\sc i} (5167.3)\\
  3030015215917673088 & 10.48 & Mg {\sc ii} (4481.1), Fe {\sc ii} (4549.5, 5169.0), Ti {\sc ii} (4572.0), Mg {\sc i} (5167.3)\\
  3029919592765618304 & 10.49 & Mg {\sc ii} (4481.1), Fe {\sc ii} (4549.5, 5169.0), Mg {\sc i} (5167.3, 5172.7)\\
  3030069263785446400 & 10.51 & Mg {\sc ii} (4481.1), Fe {\sc ii} (4549.5, 4923.9, 5169.0), Mg {\sc i} (5167.3, 5172.7), Fe {\sc i} (4918--4958)\\
  3030313802042017536 & 10.57 & Mg {\sc ii} (4481.1), Fe {\sc ii} (4549.5, 5169.0), Mg {\sc i} (5167.3, 5172.7)\\
  3030015662586533888 & 10.68 & Mg {\sc ii} (4481.1), Fe {\sc ii} (4549.5, 5169.0), Mg {\sc i} (5167.3, 5172.7)\\
  3030026588989698048 & 10.70 & Mg {\sc ii} (4481.1), Fe {\sc ii} (4549.5, 5169.0), Mg {\sc i} (5167.3, 5172.7)\\
  3030028684933623808 & 10.76 & Mg {\sc ii} (4481.1), Fe {\sc ii} (4549.5, 5169.0), Mg {\sc i} (5167.3, 5172.7)\\
  3030016109262918016 & 11.09 & Mg {\sc ii} (4481.1), Fe {\sc ii} (4549.5, 5169.0), Mg {\sc i} (5167.3, 5172.7)\\
  3030022152277677440 & 11.26 & Mg {\sc ii} (4481.1), Fe {\sc ii} (4549.5, 5169.0), Mg {\sc i} (5167.3, 5172.7)\\
  3030014013325413248 & 11.37 & Mg {\sc ii} (4481.1), Fe {\sc i} (4957.3, 4957.6), Fe {\sc ii} (5169.0), Mg {\sc i} (5167.3, 5172.7)\\
  3030026138007337088 & 11.49 & Mg {\sc ii} (4481.1), Fe {\sc ii} (4549.5, 5169.0), Mg {\sc i} (5167.3, 5172.7)\\
  3030025661276778880 & 11.63 & Fe {\sc ii} (4549.5, 5169.0), Mg {\sc i} (5167.3, 5172.7, 5183.6)\\
\hline\end{tabular}
\begin{tablenotes}
\footnotesize
\item[a] ID in \textit{Gaia} EDR3
\item[b] The wavelength values in the brackets are in units of
  \text{\AA}, from \url{http://kurucz.harvard.edu}
  \citep{2005MSAIS...8...86K, 2011CaJPh..89..417K,
    2018ASPC..515...47K}. To fit each spectrum, not all but more than
  two lines listed for the corresponding star were fitted.
\end{tablenotes}
\end{threeparttable}
\end{table*}

\begin{table*}
\scriptsize
	\centering
	\caption{RV Measurement Results\label{tab:fitting_result}}
    \begin{threeparttable}
\begin{tabular}{|c|c|c|c|c|c|c|c|c|}

\hline\hline
  \multicolumn{1}{|c|}{\textit{Gaia} ID\tnote{a}} &
  \multicolumn{1}{c|}{$v\sin{i}$($\unit{km\,s^{-1}}$)\tnote{b}} &
  \multicolumn{1}{c|}{$\mathrm{RV_1}$($\unit{km\,s^{-1}}$)\tnote{c}} &
  \multicolumn{1}{c|}{$\mathrm{RV_2}$($\unit{km\,s^{-1}}$)\tnote{c}} &
  \multicolumn{1}{c|}{$\mathrm{RV_3}$($\unit{km\,s^{-1}}$)\tnote{c}} &
  \multicolumn{1}{c|}{$\mathrm{RV\ error}$($\unit{km\,s^{-1}}$)\tnote{d}} &
  \multicolumn{1}{c|}{RV variation($\unit{km\,s^{-1}}$)} &
  \multicolumn{1}{c|}{RV variation error($\unit{km\,s^{-1}}$)\tnote{e}} &
  \multicolumn{1}{c|}{Obs number\tnote{f}} \\
\hline
  3030259479295155072 & $57\pm1.7$ & -1 & 85 & 86 &$<2$& 87 & $<2.6$ & 3\\
  3030027447983067008 & $102\pm 2.9$ & 34 & 34 & 36 & $<4$ & 2 & $<4.0$ & 3\\
  3028387801268979584 & $40\pm 1.7$ & 43 & 54 & 39 &$<2$& 15 & $<2.8$ & 3\\
  3030231785345188608\tnote{*} & $88\pm 3.5$ & 33 & 36 & -- &$<3$& 3 & $<3.8$ & 2\\
  3030030746517945600 & $45\pm 1.7$ & 34 & 34 & 35 &$<2$& 1 & $<2.0$ & 3\\
  3030250751920573696\tnote{*} & $35\pm 1.7$ & 35 & 35 & 36 &$<2$& 1 & $<2.0$ & 3\\
  3029232707231846784 & $100\pm 3.5$ & 33 & 33 & -- &$<3$& 0 & $<2.6$ & 2\\
  3030038546178432256 & $70\pm2.1 $ & 35 & 35 & -- &$<2$& 0 & $<1.7$ & 2\\
  3030298374519750912\tnote{*} & $20\pm 1.7$ & 34 & 35 & 34 &$<1$& 1 & $<1.2$ & 3\\
  3030015215917673088 & $30\pm 1.7$ & 34 & 36 & 32 &$<2$& 4 & $<2.7$ & 3\\
  3029919592765618304 & $118\pm 2.9$ & 36 & 34 & 33 &$<5$& 3 & $<5.1$ & 3\\
  3030069263785446400 & $30\pm 2.1$ & 37 & 31 & -- &$<1$& 6 & $<1.4$ & 2\\
  3030313802042017536 & $70\pm1.7$ & 33 & 34 & 35 &$<2$& 2 & $<2.3$ & 3\\
  3030015662586533888 & $37\pm 1.7$ & 34 & 34 & 34 &$<2$& 0 & $<1.8$ & 3\\
  3030026588989698048 & $65\pm1.7$ & 34 & 33 & 33 &$<3$& 1 & $<2.9$ & 3\\
  3030028684933623808\tnote{*} & $25\pm 3$ & -- & 32 & -- &$<2$& -- & -- & 1\\
  3030016109262918016 & $65\pm2.1$ & -- & 31 & 32 &$<2$& 1 & $ <2.0$ & 2\\
  3030022152277677440\tnote{*} & $65\pm2.1$ & -- & 33 & 35 &$ <3$& 2 & $<3.3$ & 2\\
  3030014013325413248 & $80\pm 5$ & -- & 34 & -- &$<2$& -- & -- & 1\\
  3030026138007337088 & $85\pm 5$ & -- & 33 & -- &$<4$& -- & -- & 1\\
  3030025661276778880 & $25\pm 1.7$ & 36 & 36 & 37 &$<2$& 1 & $<2.0$ & 3\\
\hline\end{tabular}

\begin{tablenotes}
\footnotesize
\item[a] ID in \textit{Gaia} EDR3;
\item[b] Measured mean $v\sin{i}$ values and their $2\sigma$ errors;
\item[c] $\mathrm{RV_1}$, $\mathrm{RV_2}$ and $\mathrm{RV_3}$ are the
  RVs observed in November 2020, November 2021 and December 2021,
  respectively, except for \textit{Gaia} ID 3030025661276778880, for
  which data were obtained on 26 November 2021, 27 December 2021 and
  28 December 2021, respectively;
\item[d] $2\sigma$ uncertainty of the RV measurement; 
\item[e] $2\sigma$ uncertainty of the RV variations;
\item[f] Number of observations of the corresponding star;
\item[*] Stars classified as rMS stars by \citetalias{2022ApJ...938...42H}.
\end{tablenotes}
\end{threeparttable}
\end{table*}

\section{Main results}\label{sec:result}

In Table \ref{tab:fitting_result}, we present the measured RV values
as well as the ranges of their $2\sigma$ uncertainties. The mean
values of the measured $v\sin i$ for multiple observations for each
star with their $2\sigma$ uncertainties are also listed. We confirm
that the variations in the observed $v\sin i$ among multiple
observations for each star are within the $v\sin i$ step
($\unit[5]{km\,s^{-1}}$) of our spectroscopic fits. We note that the
measured $v\sin i$ of \textit{Gaia} ID 302991959276561830 exceeds the
upper limit of $v\sin i$ we adopted for selecting slowly rotating
targets for RV measurements, based on
\citetalias{2022ApJ...938...42H}. This may be caused by the different
selection of absorption lines fitted in this paper compared with that
of \citetalias{2022ApJ...938...42H} for this particular target. Since
its $v\sin i$ is close to the $v\sin i$ upper limit for selecting
slowly rotating targets, we include this star in the discussion below.

\begin{figure*}
    \includegraphics[width=0.5\textwidth]{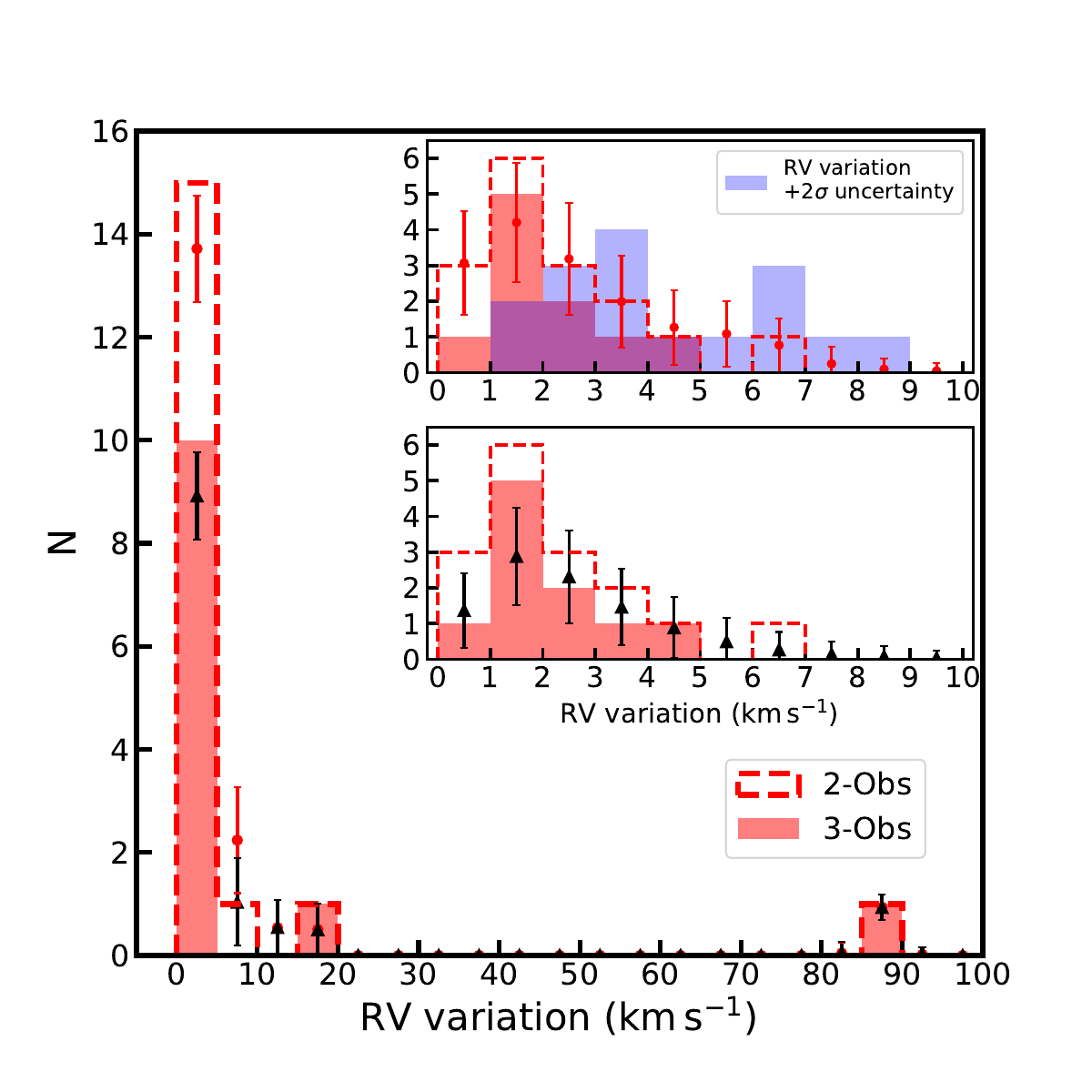}  \\
    \caption{Number distributions of the measured RV variations of the
      2-Obs (open and dashed symbols) and 3-Obs stars (filled
      symbols). The insets show the number distributions of the RV
      variations for the non-RV-variable candidates in these two
      populations, where the light blue bars in the top-right inset show the distribution of
      their RV variations plus their $2\sigma$ uncertainties (see
      Table \ref{tab:fitting_result}). The error bar shows the average value and the standard deviation of the number in each RV variation bin for the 10,000 sets of pseudo RV variations (see the text), for the 2-Obs stars (red dots) and 3-Obs stars (black triangles).}\label{fig:RV_var}
\end{figure*}

The RV variations for each star, defined as the largest absolute difference among
the RVs of a given star at different epochs, are also listed in Table
\ref{tab:fitting_result}. The number distributions of the RV
variations of the 2-Obs and 3-Obs stars are plotted in Figure
\ref{fig:RV_var}. To estimate the uncertainty in the RV variations, we generated 10,000 pseudo-RV values for each observed RV of every star based on the observed RV and its uncertainty, then obtained 10,000 RV variations for each star. Then, the uncertainty of the RV variations of a star is the standard deviation of its pseudo RV variations. In Table
\ref{tab:fitting_result}, we show the 2$\sigma$ uncertainty in the RV variations. Using this method, we also obtained 10,000 sets of pseudo RV variations for our targets. Based on the 10,000 sets of pseudo RV variations, we calculated the average number and the standard deviation of the number of stars in each RV variation bin, then derived an average distribution of RV variations for the 10,000 pseudo RV variation sets. The average distributions of the pseudo RV variation sets for different populations are shown in Figure \ref{fig:RV_var}. Using these average distributions, we determined the influence of the RV measurement errors on the distribution of observed RV variations. 

Most (16 out of 18) 2-Obs stars show small RV variations $\leq\unit[6]{km\,s^{-1}}$. Only two stars (\textit{Gaia}
ID 3028387801268979584 and 3030259479295155072) exhibit large RV
variations $\geq \unit[15]{km\,s^{-1}}$. \textit{Gaia} ID
3028387801268979584, whose RV variation is $\unit[15]{km\,s^{-1}}$,
shows a RV difference $\geq\unit[10]{km\,s^{-1}}$ between subsequent
observations. This star was found to be a double-lined spectroscopic
binary system by \citetalias{2022ApJ...938...42H}. \textit{Gaia} ID
3030259479295155072 exhibits the largest RV variation among our
targets, $\unit[87]{km\,s^{-1}}$, between observations obtained in
November 2020 and December 2021. We note that the RV difference for
this star between the observations of November 2021 and December 2021
is as small as $\unit[1]{km\,s^{-1}}$. This may result from the
similar orbital phases at the time that both observations were
taken. We refer to these two stars which have RV variations $\geq
\unit[15]{km\,s^{-1}}$ as RV variable candidates. Their positions in
the CMD are shown in Figure~\ref{fig:RV_samples}. The other 16 2-Obs
stars and 10 3-Obs stars display relatively small RV variations ($\leq
\unit[6]{km\,s^{-1}}$) compared with those of the RV variable
candidates. In the top-right inset of Figure \ref{fig:RV_var}, we plot the
distribution of their RV variations and that of their RV variations
plus $2\sigma$ uncertainties, which can be adopted as upper limits to
their RV variations. Figure \ref{fig:RV_var} shows that all upper limits to the RV variations are smaller than
$\unit[10]{km\,s^{-1}}$. We refer to these 16 2-Obs stars
and 10 3-Obs stars, whose upper limits to the RV variations $<
\unit[10]{km\,s^{-1}}$, as our non-RV-variable candidates. They
comprise 89 per cent (16/18) and 83 per cent (10/12) of the 2-Obs and
3-Obs stars, and 87 per cent (13/15) and 82 per cent (9/11) of the
2-Obs-shade and 3-Obs-shade stars, respectively. The fractions of the 2-Obs and 3-Obs stars showing RV variations $\leq\unit[6]{km\,s^{-1}}$ are $82\pm 5$ per cent and $78\pm 6$ per cent in the sets of pseudo RV variations, respectively.

In \citetalias{2022ApJ...938...42H}, we estimated the mean
  synchronisation time-scales, $\tau_{\mathrm{syn}}$, of 10 split-MS
  stars with $v\sin{i}<\unit[100]{km\,s^{-1}}$ using the theoretical
  equation (44) of \cite{2002MNRAS.329..897H}. This equation is based on the dynamical tides theory proposed by \cite{1975A&A....41..329Z,1977A&A....57..383Z} \footnote{ Based on \cite{1975A&A....41..329Z,1977A&A....57..383Z}, the
      factor $52^{5/3}$ in this equation should be $5\times 2^{5/3}$
      instead. Therefore, $\tau_{\mathrm{syn}}$ derived by
      \citetalias{2022ApJ...938...42H} are incorrect. We published an erratum \citep{2023ApJ...952..172H} for \citetalias{2022ApJ...938...42H} showing revised $\tau_{\mathrm{syn}}$ values as well as relevant discussion and figure, newly derived using the corrected equation (44) of
      \cite{2002MNRAS.329..897H}. This paper is based on the revised
      results.}. In \citetalias{2022ApJ...938...42H}, the correlation between $\tau_{\mathrm{syn}}$ and
  the fractional binary separation, $a/R$, was derived, where $a$ is
  the separation between the primary and the companion star and $R$
  the radius of the primary star \citepalias[see figure 12 of
  ][]{2022ApJ...938...42H}. In this work, we use $a$ and $R$ to represent the same quantities as in \citetalias{2022ApJ...938...42H}.
  In \citetalias{2022ApJ...938...42H}, we
  found that $\tau_{\mathrm{syn}}$ increased dramatically with
  increasing $a/R$. To fully tidally lock the binaries on time-scale on the
  order of the cluster age, a mean $a/R<4.5$ was required
  \citepalias{2022ApJ...938...42H}. This indicates that large RV
  variations are expected in time-domain observations if our targets
  are close binaries which have been tidally locked on time-scales
  similar to the cluster age ($\sim \unit[90]{Myr}$).

To compare the observed RV variations with those expected for binaries that have been fully or partially synchronised according to the dynamical tides theory, we modelled the RV variations when two-
and three-epoch observations are taken for the 2-Obs and
3-Obs stars, respectively, using the $a/R$ range derived by \citetalias{2022ApJ...938...42H}. We first modelled assuming that their rotation rates were fully tidally braked
within $\unit[90]{Myr}$, referring to the cases of two- and three-epoch observations as Case 2-Obs and Case 3-Obs, respectively. We assumed circularised orbits for the
tidally locked binaries. As the observed RV of a star is the
projection of its velocity along its orbit, the RV would be correlated
with its $v\sin i$ instead of the equatorial rotation rate if the star
were tidally locked. The radial velocity of the binary component,
$\mathrm{RV_{obs}}$, of a tidally locked binary can thus be expressed
as
\begin{equation}\label{RV_syn}
\mathrm{RV_{obs}}=\mathrm{RV_c}+\frac{a}{R} v\sin{i} \frac{q}{1+q}
\cos{\phi_{t}},
\end{equation}
where $\mathrm{RV_c}$ is the centroid RV, $q$ the mass ratio of the
primary to the companion and $\phi_{t}$ the phase of the binary orbit
at the observation time, \textit{t}. In our models, we adopted the
average RV ($\sim \unit[36]{km\,s^{-1}}$) of 57 NGC 2422 FGK-type
stars measured by \cite{2018MNRAS.475.1609B} as $\mathrm{RV_c}$. The
input $v\sin i$ of the stars are their measured mean $v\sin i$, listed
in Table \ref{tab:fitting_result}. According to
\citetalias{2022ApJ...938...42H}, the slowly rotating stars along the
split MS of NGC 2422 should have intermediate mass ratios between 0.3
and 0.6 if they are photometric binaries. In each run, an intermediate
$q$, uniformly distributed between 0.3 and 0.6, was thus applied for
each star. High-mass-ratio close binaries may exhibit small RV
variations because the shift in the absorption lines might be reduced
owing to the comparable contribution of flux from both
components. Since our targets have intermediate mass ratios if they
are unresolved binaries, this effect would not be significant. The $a/R$ values adopted are uniformly distributed between 3 and 4.5
  so as to synchronise the stellar rotation rates and the orbits of
  the slow rotators on time-scales shorter than the cluster age
  \citepalias{2022ApJ...938...42H}. In each run of the Case 2-Obs and
Case 3-Obs, every star was observed twice and three times,
corresponding to two and three randomly generated $\phi_{t}$,
uniformly distributed between 0 and $2\pi$, respectively. In each
  run, we also calculated the orbital periods, $T_{\mathrm{orb}}$, of
  the synthetic binaries using Kepler's Third Law. Then the equatorial
  rotation rates of the primary stars, $v_{\mathrm{eq}}$, were
  estimated using $2\pi R/T_{\mathrm{orb}}$, where $R$ was estimated
  from the empirical relation $R\approx 1.33\times{M}^{0.555}$ for
  stars with $M>\unit[1.66]{M_\odot}$, where $M$ is given in units of
  $\mathrm{M_\odot}$ and $R$ in units of
  $\mathrm{R_\odot}$ \citep{1991Ap&SS.181..313D}. We calculated
  $v\sin{i}/v_{\mathrm{eq}}$ for every star. For tidally locked
  binaries, $v\sin{i}/v_{\mathrm{eq}}$ should be smaller than unity.
We repeated this procedure 10,000 times for each case and
recorded the expected RV
variations and $v\sin{i}/v_{\mathrm{eq}}$. For Case 2-Obs and Case 3-Obs, 70 per cent and 78 per
  cent, respectively, of the $v\sin{i}/v_{\mathrm{eq}}$ ratios in the
  10,000 runs were less than unity. This implies that the set of $a$
  and $q$ of most mock binaries were consistent with the assumption
  that they are tidally locked and show the measured $v\sin{i}$
  values.

 We additionally modelled binaries with $6\leq a/R\leq 8$ to
  explore the RV variations in case these stars are partially
  synchronised. This $a/R$ range corresponds to $\tau_{\mathrm{syn}}$
  ranging from $\sim \unit[1]{Gyr}$ to $\sim \unit[10]{Gyr}$
  \citepalias[][see their figure 12]{2022ApJ...938...42H}. Based on
  the models of \cite{2017NatAs...1E.186D}, stars whose rotation rates
  are reduced significantly but which are not tidally locked could also form a blue sequence. We refer to the models for $6\leq a/R\leq
  8$ as Case 2-Obs-p and Case 3-Obs-p for our two- and three-epoch
  observations, respectively. Then the RV variations of stars whose
  rotation rates are reduced significantly but which are not tidally
  locked within $\unit[90]{Myr}$ should be located within the ranges
  of the Case 2-Obs (Case 3-Obs) and Case 2-Obs-p (Case 3-Obs-p) for
  the two-epoch (three-epoch) observations. For Case 2-Obs-p and Case
  3-Obs-p, equation (\ref{RV_syn}) cannot be used, since the periods
  of stellar rotation and the binary orbits are not synchronised.
  $\mathrm{RV_{obs}}$ is instead expressed as
\begin{equation}\label{RV_Nsyn}
\mathrm{RV_{obs}}=\mathrm{RV_c}+v_{\mathrm{orb}} \sin{j} \cos{\phi_{t}},
\end{equation}
where $v_{\mathrm{orb}}$ is the velocity of the primary star along its
binary orbit and $j$ is the inclination angle of the binary orbital
rotation axis. Using Kepler's Third Law, we calculated the orbital
periods, $T_{\mathrm{orb}}$, for the mock binaries with $a/R$
uniformly distributed between 6 and 8, and $q$ uniformly distributed
between 0.3 and 0.6, assuming circularised binary orbits. The
$v_{\mathrm{orb}}$ values were then derived using $2\pi qa/(1+q) T_{\mathrm{orb}}$. The orientations of the orbital rotation axes
were assumed to be stochastic in three-dimensional space, which was
modelled using a uniform distribution of $\cos{j}$ between 0 and 1
\citep{2019NatAs...3...76L}. The set of $\phi_{t}$ values was the same
for Case 2-Obs and Case 3-Obs. For each case, we repeated this process
10,000 times and recorded the synthetic RV variations. To test whether
the synthetic binaries were indeed not tidally locked, we estimated
their stellar rotation periods, $T_{\mathrm{rot}}$, using $2\pi
R/v\sin{i}$ multiplied by a distribution of $\sin{i}$ generated like
$\sin{j}$. In the 10,000 runs for Case 2-Obs-p and Case 3-Obs-p, 93
per cent and 90 per cent of the values of
$T_\mathrm{rot}/T_{\mathrm{orb}}$ are less than unity, indicating that
they are not tidally locked.

Figure~\ref{fig:synthetic_RV} shows the density and cumulative
distributions of the synthetic and observed RV variations for the
  Case 2-Obs, Case 3-Obs, Case 2-Obs-p and Case 3-Obs-p. In these four
cases, it is evident that the synthetic RV variations exhibit much
larger dispersions than the observations. For Case 2-Obs and Case
3-Obs, 82 and 96 per cent of the synthetic RV variations are larger
than $\unit[10]{km\,s^{-1}}$, with mean synthetic RV variations of
$\unit[55]{km\,s^{-1}}$ and  $\unit[76]{km\,s^{-1}}$,
respectively. For Case 2-Obs-p and Case 3-Obs-p, 78 and 95 per
  cent of synthetic RV variations are larger than
  $\unit[10]{km\,s^{-1}}$, with mean synthetic RV variations of
  $\unit[40]{km\,s^{-1}}$ and $\unit[61]{km\,s^{-1}}$,
  respectively. The synthetic results indicate a high probability, 82
  per cent for Case 2-Obs, 96 per cent for Case 3-Obs, 78 per cent for
  Case 2-Obs-p and 95 per cent for Case 3-Obs-p, to detect RV
  variations $>\unit[10]{km\,s^{-1}}$ if the stars have corresponding
  fractional binary separations. Additionally, the chance to detect
more stars with RV variation $\leq \unit[10]{km\,s^{-1}}$ is
small. For Case 2-Obs (Case 2-Obs-p), only 3 (17) out of the
  10,000 runs have more than half of the stars showing RV variations
  $\leq\unit[10]{km\,s^{-1}}$. Not a single run among the 10,000 runs
  for Case 3-Obs or Case 3-Obs-p has more than half of the stars
  showing RV variations $\leq\unit[10]{km\,s^{-1}}$. However, our
measurement results reveal that 89 per cent of 2-Obs and 83 per cent
of 3-Obs stars are non-RV-variable candidates with RV variations
{$\leq \unit[10]{km\,s^{-1}}$. For the 2-Obs-shade and 3-Obs-shade
stars, the non-RV-variable candidates make up 87 and 82 per cent of
their populations, respectively. In the 10,000 sets of pseudo RV variations, the fractions of the 2-Obs and 3-Obs stars showing RV variations $\leq \unit[10]{km\,s^{-1}}$ are $89\pm 1$ per cent and $83\pm 2$ per cent, respectively. Only one star (\textit{Gaia} ID
3030259479295155072) among the RV variable candidates which have
three-epoch observations shows a comparable RV variation to the mean
synthetic RV variations of the Case 3-Obs and the Case 3-Obs-p
models. The distributions of the observed RV variations are different
from those of the corresponding synthetic RV variations for the 2-Obs
and 3-Obs stars. We applied Anderson--Darling tests for $k$ samples to
explore whether the observed and synthetic RV variations are drawn
from the same distribution. The tests for the four cases
  modelled all reported significance levels $p<0.001$, thus ruling
out the hypothesis that the observed and synthetic RV variations come
from the same distribution.

\begin{figure*}
\begin{tabular}{cc}
    \includegraphics[width=0.4\textwidth]{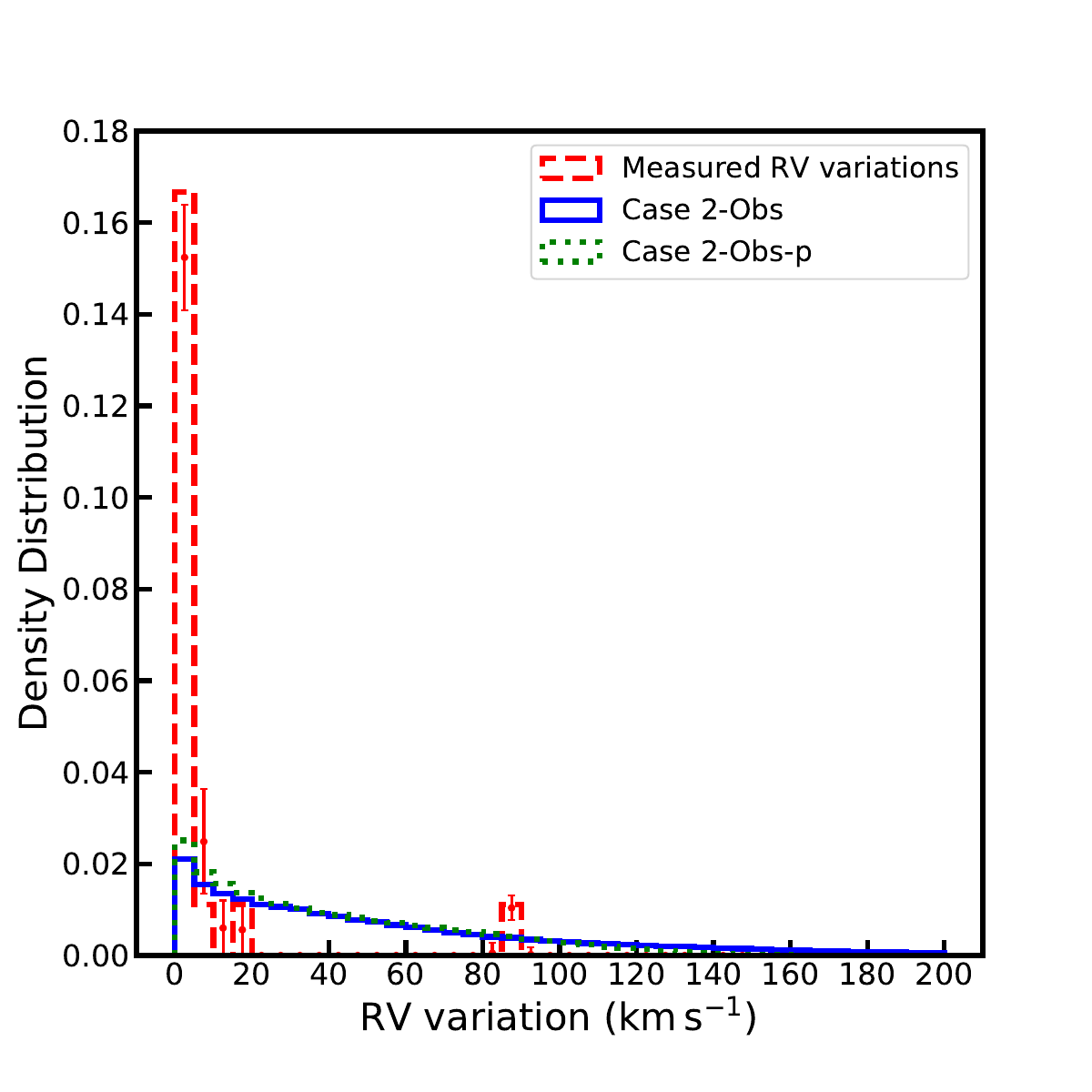}  & \includegraphics[width=0.4\textwidth]{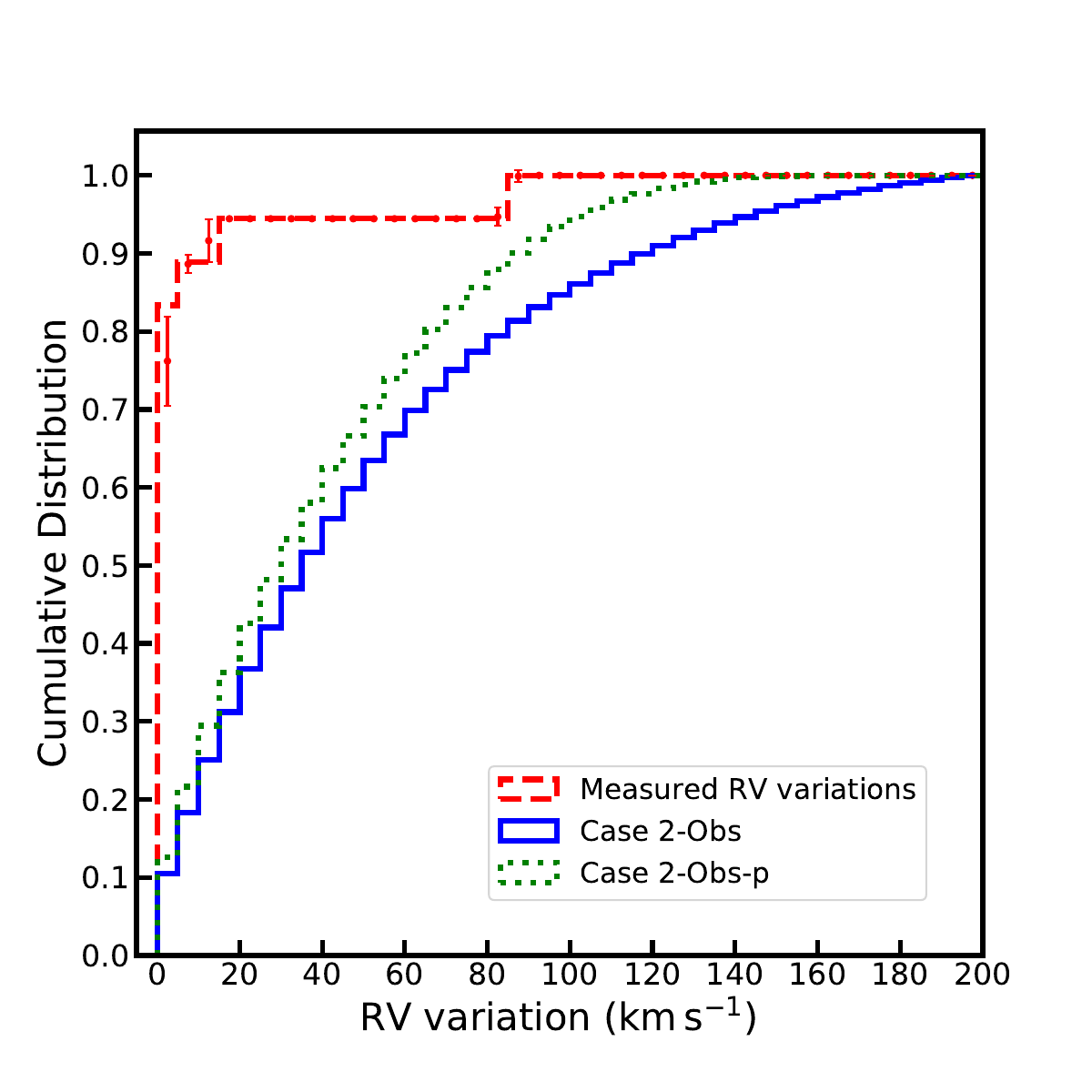}  \\
    \includegraphics[width=0.4\textwidth]{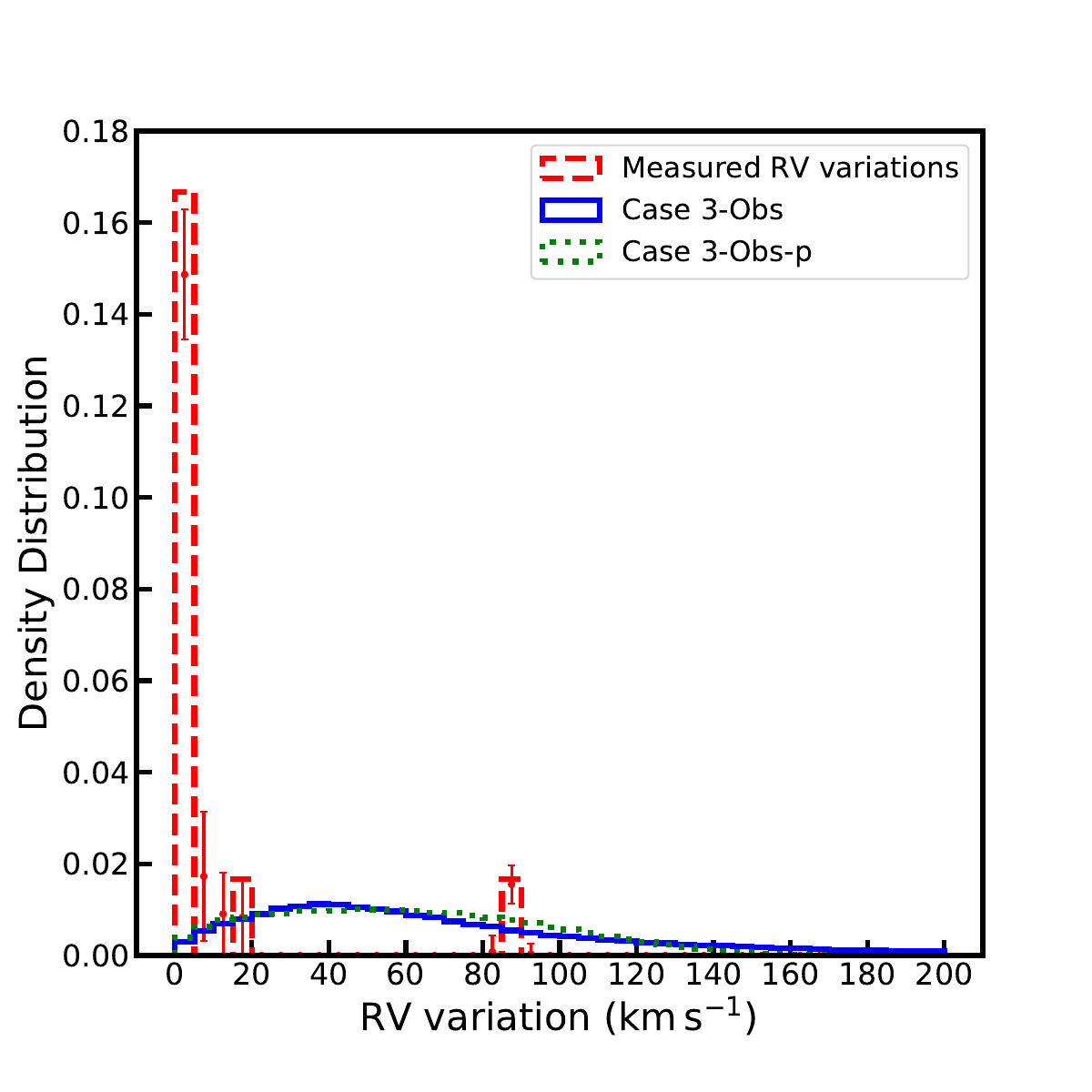} &
    \includegraphics[width=0.4\textwidth]{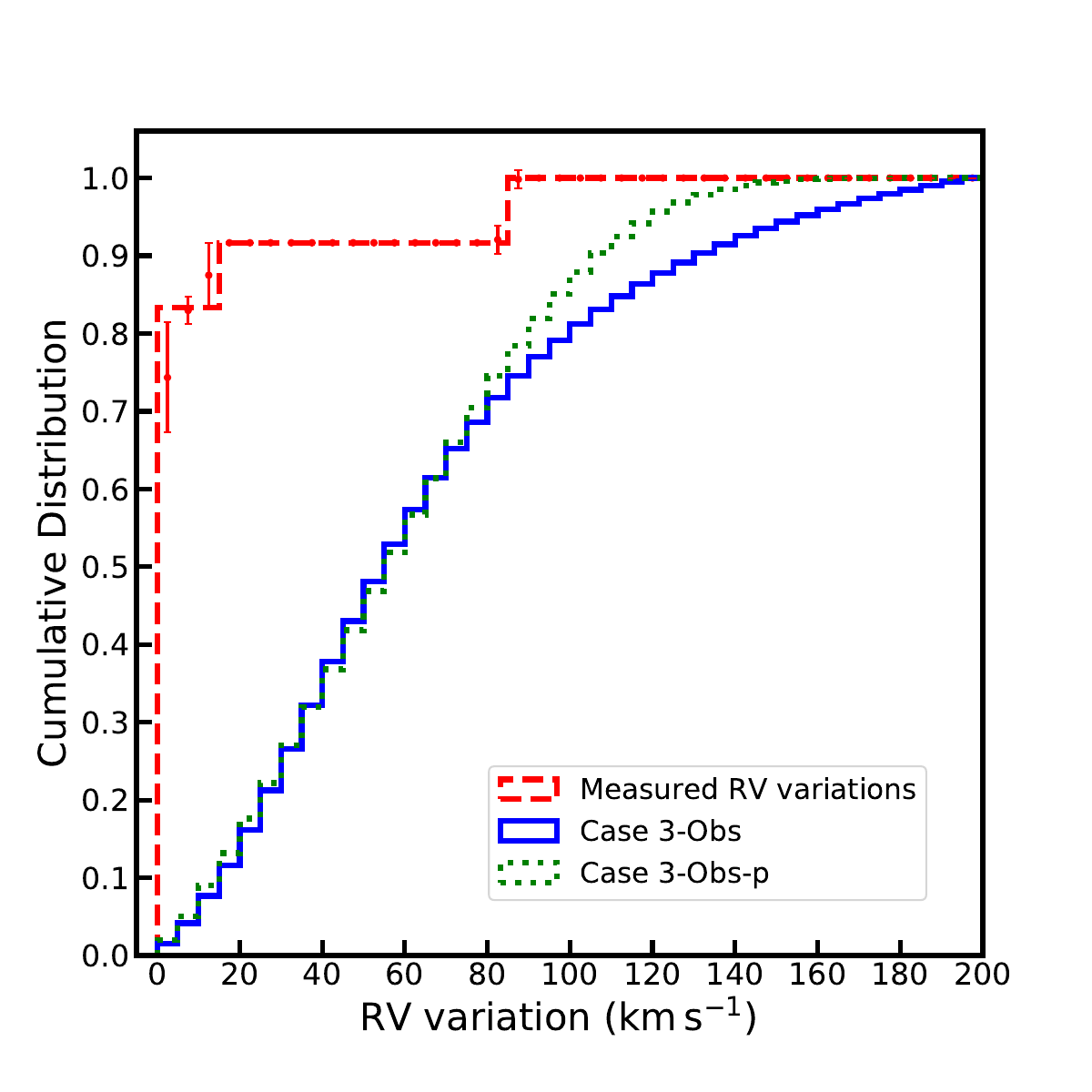} \\
\end{tabular}
	\caption{Density (left) and cumulative (right) distributions
          of the observed and synthetic RV variations. The observed RV
          variations are shown as red dashed lines in all panels. The
          two top panels show the distributions of the observed and
          synthetic RV variations for the 2-Obs stars. The
            synthetic RV variations for Case 2-Obs (Case 3-Obs) and
            Case 2-Obs-p (Case 3-Obs-p) are shown as the blue solid
            and green dotted lines in the two top (bottom) panels,
            respectively. The error bar shows the average value and the standard deviation of the number density (left) and the cumulative fraction (right) of the RV variations in each bin for the 10,000 sets of the pseudo RV variations of the 2-Obs stars (top two panels) and the 3-Obs stars (bottom two panels). The bin sizes in all panels are
          $\unit[5]{km\,s^{-1}}$}\label{fig:synthetic_RV}
\end{figure*}

A large RV dispersion during a single epoch is expected if our
  targets are characterised by $a/R\leq 8$. In Figure~\ref{fig:1obs},
we show the synthetic RV dispersion at one epoch for the 21
spectroscopic targets assuming that they have been tidally locked
  within the cluster age, i.e., $3\leq a/R\leq 4.5$ (Case 1-Obs). The
  synthetic RV dispersion for $6\leq a/R\leq 8$ (Case 1-Obs-p) is also
  shown in Figure~\ref{fig:1obs}. In Case 1-Obs, the set of
  $\mathrm{RV_c}$, $v\sin i$, $a/R$ and $q$ are identical to those of
  Case 2-Obs or Case 3-Obs. For Case 1-Obs-p, the inputs of
  $\mathrm{RV_c}$, $v_{\mathrm{orb}}$ and $\sin{j}$ are the same as
  for Case 2-Obs-p or Case 3-Obs-p. $\phi_t$ was uniformly distributed
  between 0 and $2\pi$ in each run for all models. To present the RV
dispersion, we subtracted the average $\mathrm{RV_{obs}}$ of the 21
stars from their $\mathrm{RV_{obs}}$ in each run. We repeated
  this process 10,000 times for each case. To compare with the
observations, the dispersion in the observed RVs (RVs minus the mean
RV) of the 21 stars on 26 and 27 November 2021\footnote{On these two
  days, all 21 spectroscopic targets were observed once}, is plotted
in Figure~\ref{fig:1obs}. The $1\sigma$ deviation of the observed RV
dispersion is $\unit[11.6]{km\,s^{-1}}$; it would be
$\unit[1.5]{km\,s^{-1}}$ if the two RV variable candidates are
excluded. In fact, the average RV of the spectroscopic population
excluding the RV-variable candidates is $\unit[33.8]{km\,s^{-1}}$ at
this epoch. This is consistent with the mean RV ($\sim
\unit[36]{km\,s^{-1}}$) as for the 57 FGK-type cluster members in NGC
2422 reported by \cite{2018MNRAS.475.1609B}. The $1\sigma$ deviation
for their RVs is $\unit[1.5]{km\,s^{-1}}$; the largest difference is
$\unit[2.2]{km\,s^{-1}}$ between their RV values and their mean
RV. We also generated 10,000 sets of pseudo RV dispersion for the 21 stars based on their measured RVs and the RV uncertainty. Their average distribution is plotted in Figure \ref{fig:synthetic_RV}. This process was also repeated for the 21 stars excluding the RV-variable candidates. The mean value of the standard deviation of the 10,000 sets of their pseudo RV dispersion is $\unit[2.9\pm 0.5]{km\,s^{-1}}$. The dispersion in the measured RVs of the 21 stars, excluding the
RV variable candidates, is much smaller than the synthetic RV
dispersion, whose $1\sigma$ deviation is $\unit[53.3]{km\,s^{-1}}$ for Case 1-Obs or $\unit[35.7]{km\,s^{-1}}$ for Case 1-Obs-p. Anderson--Darling tests for the observed and
  synthetic RV dispersions of the 21 samples reported significance
  levels $p<0.001$ and $p=0.001$ for Case 1-Obs and Case 1-Obs-p,
  respectively, indicating that they are not drawn from the same
  distribution, in both cases.

\begin{figure*}
\begin{tabular}{cc}
    \includegraphics[width=0.4\textwidth]{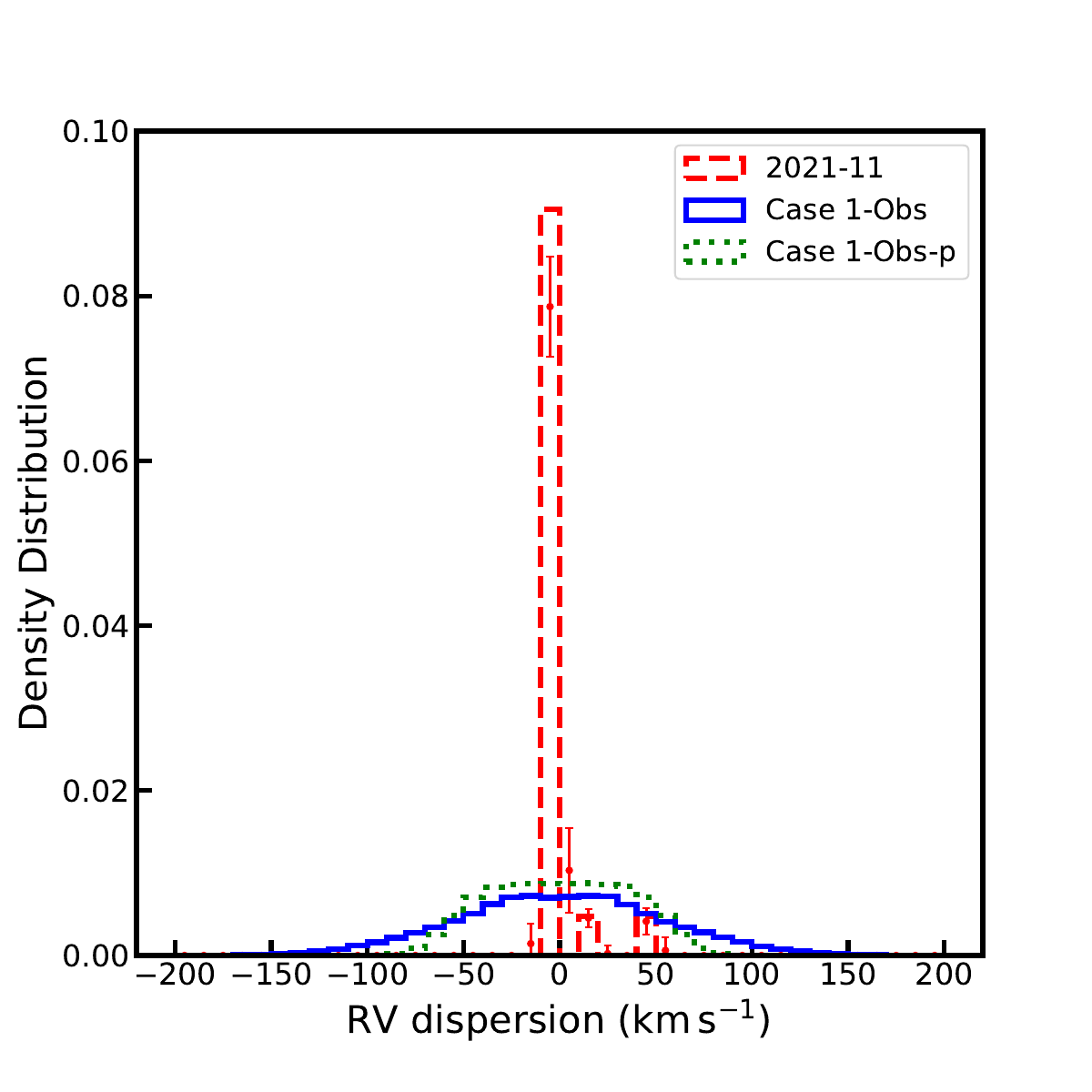}  & \includegraphics[width=0.4\textwidth]{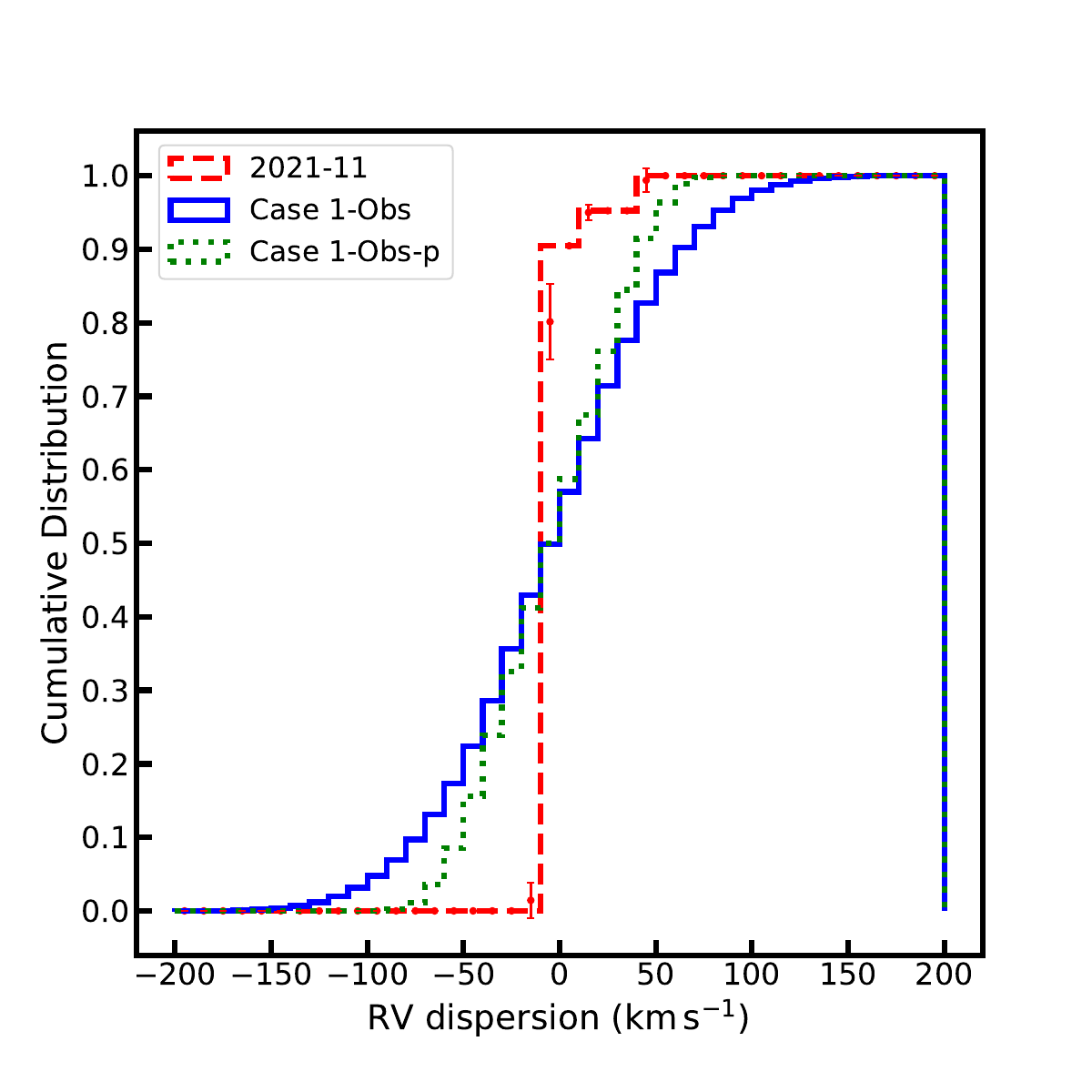}  \\
\end{tabular}
	\caption{Density (left) and cumulative (right) distributions
          of the observed RV dispersion (red dashed line) on 26 and 27
          November 2021, and the synthetic RV dispersion (blue solid
          lines) at a single epoch for the 21 targets. The error bar shows the average value and the standard deviation of the number density (left) and the cumulative fraction (right) of the RV dispersion in each bin for the 10,000 sets of the pseudo RV dispersion of the 21 targets. The bin sizes
          in all panels is $\unit[10]{km\,s^{-1}}$.}\label{fig:1obs}
\end{figure*}

\section{Discussion}\label{sec:discussion}

In Section \ref{sec:result}, we modelled the expected RV variations for binaries that were fully or partially synchronised within the cluster age, using the fractional separations derived based on the dynamical tides theory \citep{1975A&A....41..329Z,1977A&A....57..383Z,2002MNRAS.329..897H} in \citetalias{2022ApJ...938...42H}. The high fraction of stars with RV variations $\leq
\unit[10]{km\,s^{-1}}$ and the narrow dispersion of measured RVs at
the single epoch indicate that most our spectroscopic targets are not
in close binaries that can be tidally locked within the cluster age or even within $\unit[10]{Gyr}$. The non-RV-variable candidates
comprise 87 per cent (13/15) of the 2-Obs-shade and 82 per cent (9/11) of the 3-Obs-shade stars. It is thus unlikely that tidal
interactions dominate the formation of slowly rotating stars along the
split MS. We emphasise that this result is based on a comparison
with models characterised by the small $a/R$ ranges derived from the dynamical tides theory. In these models, the set of $a/R$ corresponds to $a$ separations between $\unit[0.02]{AU}$ and $\unit[0.10]{AU}$ and mean orbital periods of $\unit[1.38\pm 0.27]{days}$ (Case 2-Obs), $\unit[1.40\pm 0.28]{days}$ (Case 3-Obs), $\unit[3.52\pm 0.53]{days}$ (Case 2-Obs-p) and $\unit[3.57\pm 0.56]{days}$ (Case 3-Obs-p). The small binary separations give rise to large RV variations that are not consistent with the observed RV variation values.

However, \cite{2004ApJ...616..562A} uncovered a
  stronger effect of tidal interactions on stellar rotation than that
  predicted by the theory just described. \cite{2004ApJ...616..562A}
  explored the projected rotation rates and the orbital periods of 400
  spectroscopic or visual binaries with B0 to F0 primary stars. The
  rotation rates of primary stars in binaries with $T_{\mathrm{orb}}$
  of 4--500 days was also found to be substantially slowed down
  compared with that of single stars, presumably by tidal interactions
  \citep{2004ApJ...616..562A}. This $T_{\mathrm{orb}}$ range was also proposed by \cite{2015MNRAS.453.2637D} as the possible orbital periods for slowly rotating split MS stars if their rotation is slowed down by tidal interactions. In Figure \ref{fig:orbit v_T}, we plot
  the correlations between the amplitudes of the projected orbital
  velocities (inclination angle $j=\unit[45]{^{\circ}}$) of the
  primary stars and the binary separations $a$ for different
  intermediate mass ratios $q$, assuming circularised binary
  orbits. The primary stars in Figure \ref{fig:orbit v_T} have
  $\unit[1.8]{M_{\odot}} \leq M \leq \unit[3.6]{M_{\odot}}$, which is similar to the mass range in the shaded region of Figure \ref{fig:RV_samples}. In Figure
  \ref{fig:orbit v_T}, the $T_{\mathrm{orb}}$ of
  100--$\unit[500]{days}$ correspond to $a$ of 0.6--$\unit[1.9]{AU}$,
  which is larger than the maximum $a$ adopted in the models by a
  factor of $\geq 6$. With this $a$ range, our targets may exhibit RV variations on the order of $\unit[1]{km\,s^{-1}}$.

We modeled the RV variations for a larger $a$ range than adopted
  for the models in Section \ref{sec:result}. The models we ran were
  similar to Case 2-Obs-p and Case 3-Obs-p for the 2-Obs and 3-Obs
  stars, respectively, although the set of $\phi_{t}$ for the mock
  binaries with $T_{\mathrm{orb}} \geq \unit[10]{days}$ was
  different. Their $\phi_{t}$ included a time interval of
  $\unit[365]{days}$ between the first and the second rounds, and a
  time interval of $\unit[30]{days}$ between the second and the third
  rounds, because their $T_{\mathrm{orb}}$ are comparable with the
  time intervals between the different observation epochs. We will
  refer to the fractions of the runs in which more than half of the
  stars show RV variations $\leq \unit[10]{km\,s^{-1}}$ as
  $f_{\mathrm{half\leq 10}}$. In Figure \ref{fig:half_fraction}, we plot
  the correlations of $f_{\mathrm{half\leq 10}}$ with the corresponding
  binary separations $a$. Based on Figure \ref{fig:half_fraction}, we
  argue that the non-RV-variable candidates are not in close binaries
  with $a<\unit[0.2]{AU}$ ($T_{\mathrm{orb}}<\unit[19]{days}$), since
  $f_{\mathrm{half\leq 10}}$ is smaller than 5 per cent for this $a$
  range. Additionally, most 3-Obs stars should have
  $a>\unit[0.6]{AU}$, corresponding to $T_{\mathrm{orb}}>\sim
  \unit[100]{days}$. For binaries with $a$ of $0.6\sim \unit[1.8]{AU}$
  (grey shaded region), the probability that more than half show RV
  variations $\leq \unit[10]{km\,s^{-1}}$ is greater than 0.05,
  corresponding to $T_{\mathrm{orb}}$ of 100--$\unit[500]{days}$. The
  fluctuations in $f_{\mathrm{half\leq 10}}$ across this $a$ range are
  caused by the comparable values of $T_{\mathrm{orb}}$ to the time
  intervals between the different rounds of our observations, leading
  to a greater possibility to detect small RV variations. We modelled
  the RV variations with $a$ uniformly distributed between
  $\unit[0.6]{AU}$ and $\unit[1.8]{AU}$. The result shows that
  $f_{\mathrm{half\leq 10}}$ is 86 per cent and 35 per cent for the mock
  two- and three-epoch observations, respectively. Based on the analysis for $ f_{\mathrm{half\leq 10}}$, the possibility
  that our non-RV-variable candidates are binaries with
  $\unit[100]{days}< T_{\mathrm{orb}}<\unit[500]{days}$ cannot be
  excluded. With such long $T_{\mathrm{orb}}$, they may experience
  strong braking of stellar rotation owing to tidal interactions
  according to \cite{2004ApJ...616..562A}. In Figure \ref{fig:synthetic_RV_P100}, we plot the distributions of the RV variations of the mock binaries with $\unit[100]{days}< T_{\mathrm{orb}}<\unit[500]{days}$ as well as those of the observed values for the 3-Obs stars. It seems that the observed RV variations are smaller than the synthetic RV variations. Eighty-two per cent of the 3-Obs stars have measured RV variations $\leq \unit[5]{km\,s^{-1}}$ and they comprise all the non-RV-variable candidates in the 3-Obs stars. This fraction is $74\pm7$ per cent in the sets of the pseudo RV variations where the RV measurement uncertainty are considered. However, the fraction of the synthetic RV variations $\leq \unit[5]{km\,s^{-1}}$ is only 24 per cent for the 3-Obs stars. Nevertheless, because of the small number of the 3-Obs stars and the high possibility to detect RV variations on order of $\unit[1]{km\,s^{-1}}$ with such $T_{\mathrm{orb}}$ range, we suggest to explore the RV variability of slowly rotating massive stars in more clusters to confirm whether they are binary components with $T_{\mathrm{orb}}<\unit[500]{days}$.

\begin{figure*}
    \includegraphics[width=0.5\textwidth]{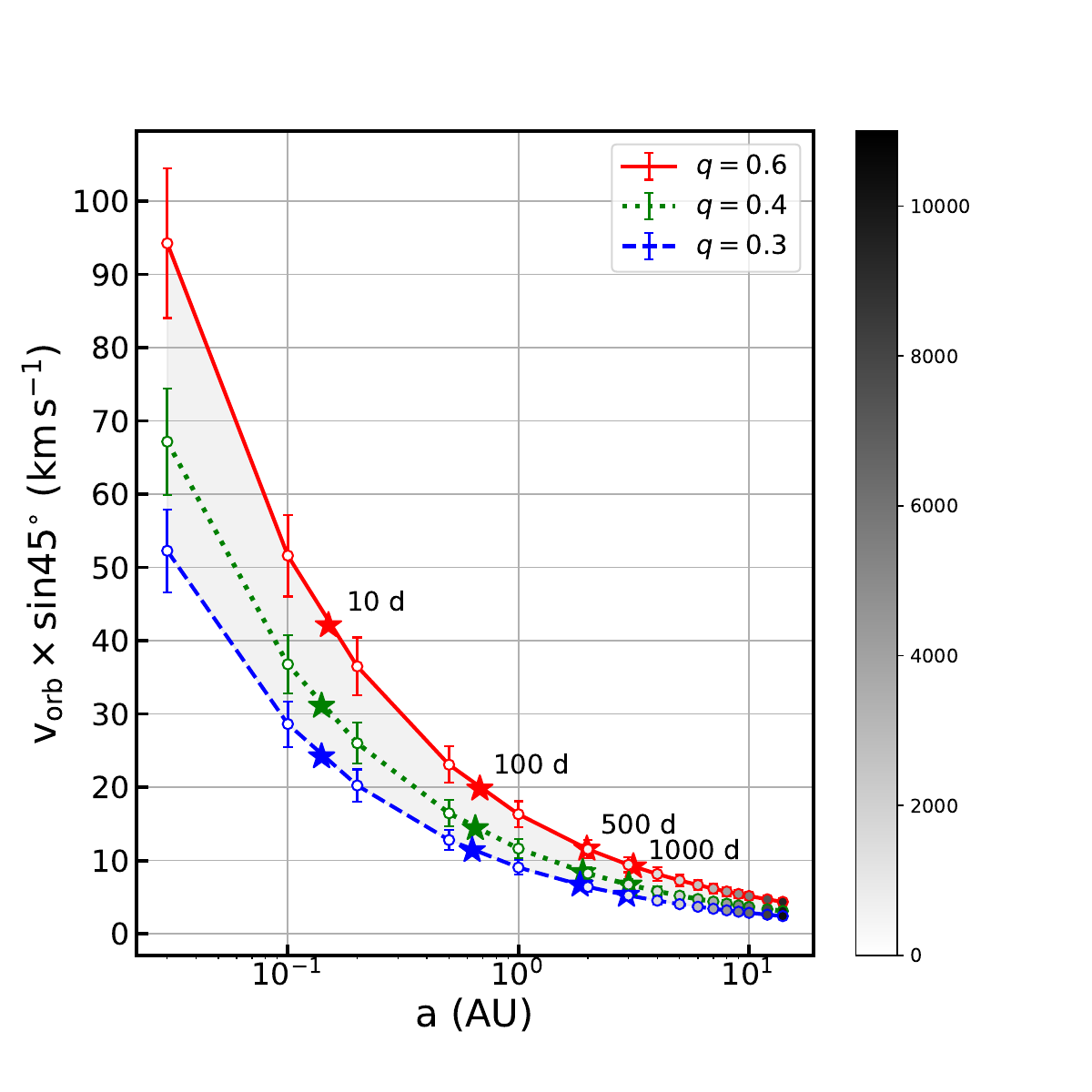} \\
    \caption{Correlation of the orbital velocity, $v_\mathrm{orb}$, of
      the primary star, multiplied by $\sin{45^{\circ}}$, with the
      binary separation, $a$, for a mass range from
      $\unit[1.8]{M_\odot}$ to $\unit[3.6]{M_\odot}$. The blue dashed,
      green dotted and red solid lines show the correlations for
      $q=0.3$, $q=0.4$ and $q=0.6$, respectively. The correlations for
      $0.3<q<0.6$ are located within the grey shaded area. The error
      bars show the standard deviations associated with the different
      primary masses. The mean values of the orbital periods, in units
      of days, are colour-coded. The star labels along each line show
      the points corresponding to the mean orbital periods as
      indicated. The correlations were calculated based on Kepler's
      Third Law, assuming circularised orbits.}\label{fig:orbit v_T}
\end{figure*}

\begin{figure*}
\begin{tabular}{cc}
    \includegraphics[width=0.5\textwidth]{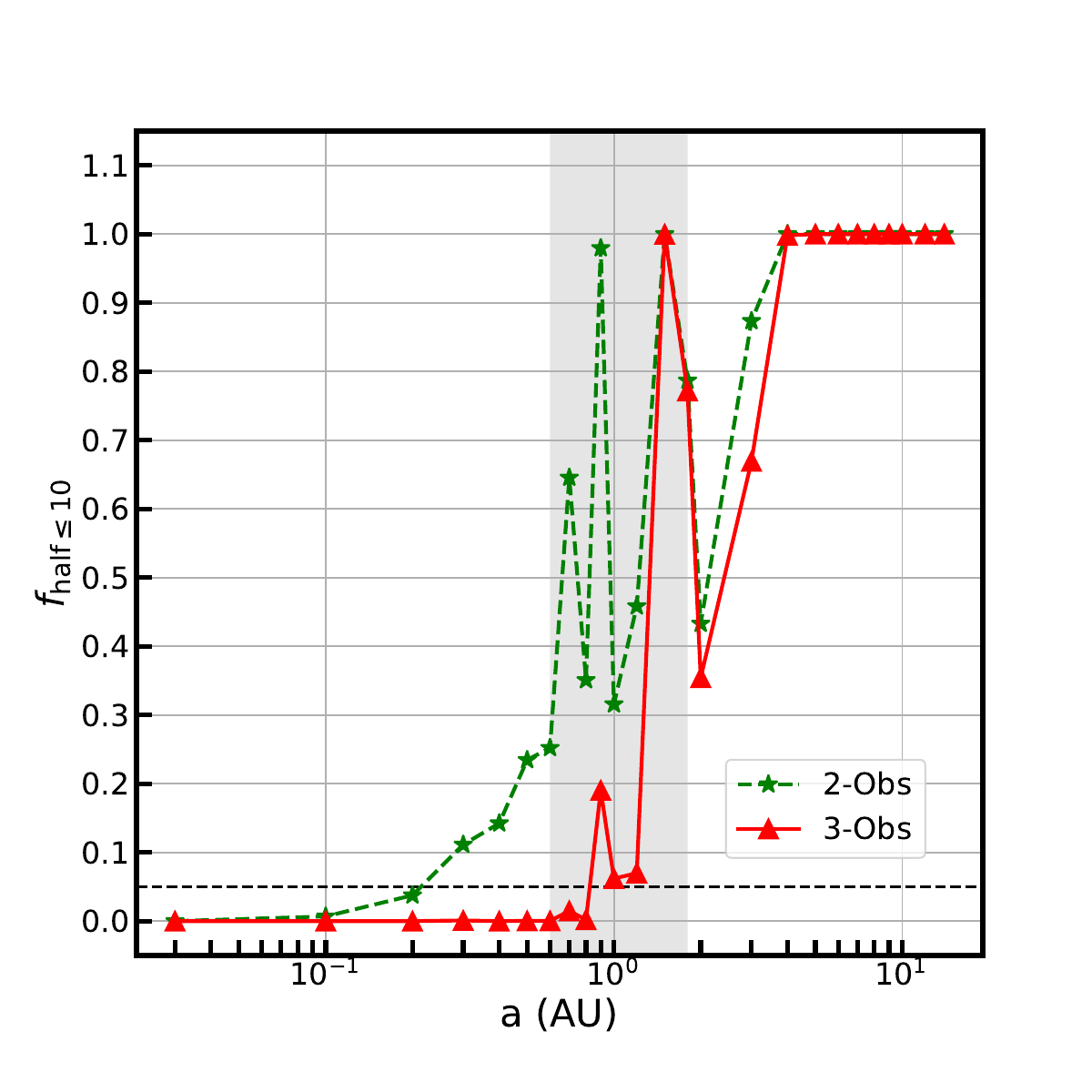} \\
\end{tabular}
	\caption{Fraction of runs in which more than half of the
            stars have RV variations $\leq \unit[10]{km\,s^{-1}}$,
            $f_{\mathrm{half\leq 10}}$, versus binary separation, $a$. The
            results for the two- and three-epoch observations are
            shown using a green dashed line with star labels and a red
            solid line with triangular labels, respectively. The grey
            shaded area shows the range
            $\unit[0.6]{AU}<a<\unit[1.8]{AU}$, corresponding to
            $\unit[100]{days}<T_{\mathrm{orb}}<\unit[500]{days}$. The
            dotted horizontal line shows the locus for a fraction of
            0.05.}\label{fig:half_fraction}
\end{figure*}

\begin{figure*}
\begin{tabular}{cc}
    \includegraphics[width=0.4\textwidth]{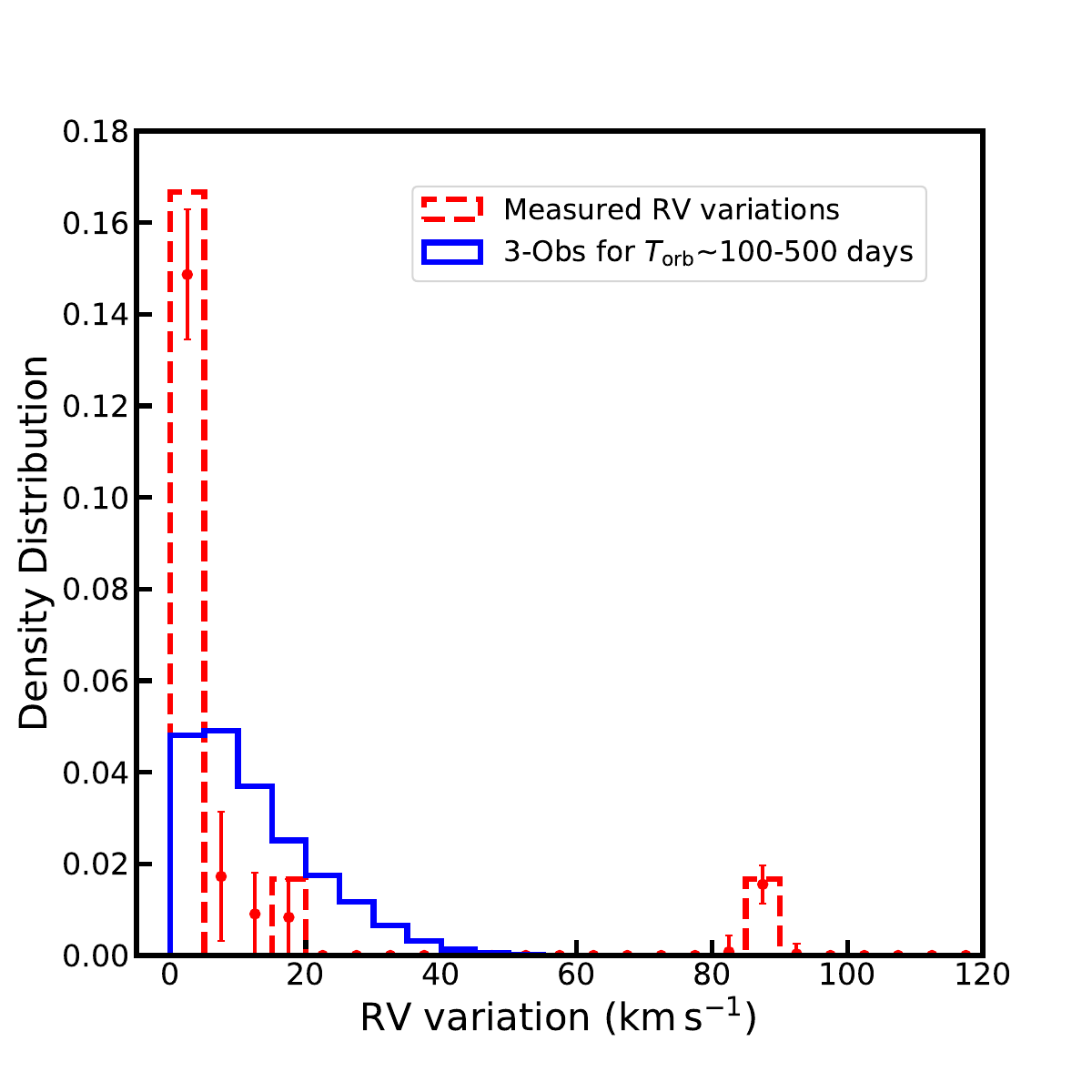} &
    \includegraphics[width=0.4\textwidth]{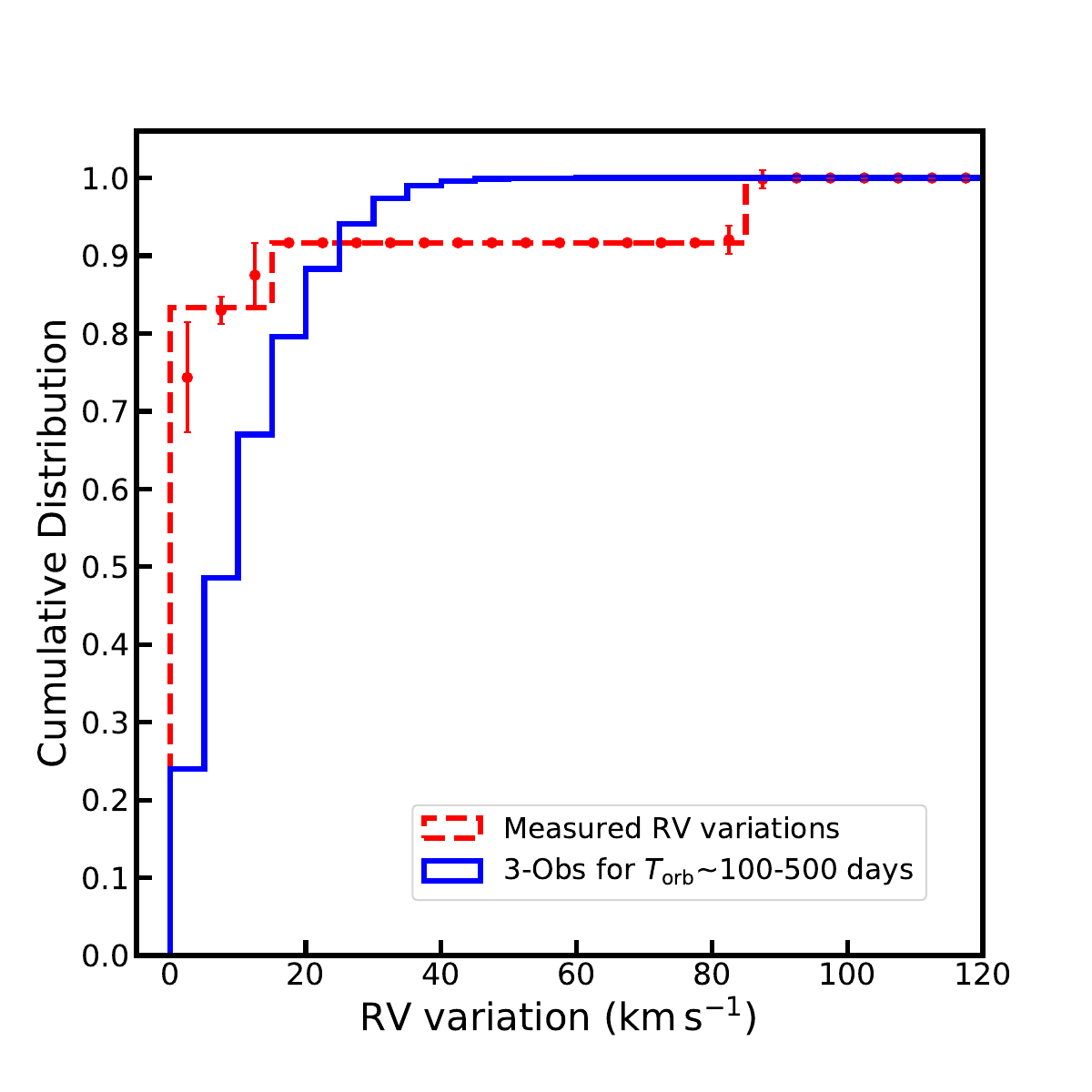} \\
\end{tabular}
	\caption{ Density (left) and cumulative (right) distributions
          of the observed and synthetic RV variations for the 3-Obs stars if they have orbital periods of 100--500 days. The distributions of the observed and synthetic RV
          variations are shown as red dashed and blue solid lines, respectively. The error bar shows the average value and the standard deviation of the number density (left) and the cumulative fraction (right) of the RV variations in each bin for the 10,000 sets of the pseudo RV variations of the 3-Obs stars. The bin sizes in both panels are
          $\unit[5]{km\,s^{-1}}$}\label{fig:synthetic_RV_P100}
\end{figure*}

Our sample stars exhibiting small RV variations might be slowly
rotating single stars or wide binary stars with
  $T_{\mathrm{orb}}>\unit[100]{days}$. According to
\cite{2020MNRAS.495.1978B}, stars with close companions would disrupt
their discs early on and tend to form rapidly rotating stars. Stars in
wide binaries or single stars with long-surviving discs would form
slow rotators, whose rotation is braked by disc locking \citep[see also][]{2012A&A...537A.120Z}. Such systems
would show RV varibility that is consistent with our observations. The
correlation between wide binaries and slow rotators was implied by
\cite{2021ApJ...912...27Y}, who found that the spatial distributions
of bMS stars were more like those of soft binaries. The binary-merger
scenario suggested by \cite{2022NatAs...6..480W} would naturally
produce slowly rotating single stars. These two scenarios need more
observational confirmation, however. For example,
\cite{2020MNRAS.495.1978B} proposed to explore the distribution of the
rotation rates of early-type stars in very young (a few Myr-old)
clusters to test whether they are bimodal. \cite{2022NatAs...6..480W}
suggested to study the slope of the mass functions of field stars or
the evolution of the stellar mass functions in star clusters, since
stellar mass functions may vary owing to binary mergers.

In \citetalias{2022ApJ...938...42H}, we detected a weak correlation
between stellar colours and rotation rates among the split-MS stars of
NGC 2422, casting some doubt on their relationship. This correlation
is caused by the appearance of a large fraction of rMS stars with
small $v\sin i$ similar to the bMS stars. In
\citetalias{2022ApJ...938...42H}, we interpreted these slowly rotating
rMS stars as photometric binaries. Their loci in the CMD are shifted
to the redder and brighter side, rendering them close to the rMS owing
to contamination by hidden low-mass companions. This explanation
cannot be ruled out. Based on \cite{2021A&A...649A...1G}, the
completeness of close source pairs drops significantly for separations
below about $\unit[0.7]{arcsec}$. This angular separation corresponds
to $\sim\unit[333]{AU}$ at the distance of NGC 2422, which is much
wider than the separations of the sample binaries discussed above. The
slowly rotating rMS stars might be unresolved binaries with
  $\unit[0.6]{AU}<a<\unit[333]{AU}$. Then, the four spectroscopic
targets in this papers classified as rMS stars by
\citetalias{2022ApJ...938...42H} with multiple observations (see Table
\ref{tab:fitting_result}) could be unresolved binaries and show small
RV variations ($\leq \unit[3]{km\,s^{-1}}$).

However, the possibility that the small $v\sin{i}$ values of the rMS
stars are caused by small inclination angles $i$ of the rotation axis
cannot be ruled out. In Figure \ref{fig:test_angle}, we show two
coeval synthetic populations derived from the SYCLIST database
\citep{2013A&A...553A..24G, 2014A&A...566A..21G}, with one population
that does not rotate whereas the other is rotating at 90 per cent of
the critical rotation rate ($\mathrm{\omega_{crit}}$). The synthetic
populations are $\sim \unit[90]{Myr}$ old, with $Z=Z_{\odot}$. The
inclination angles of both populations are randomly distributed; the
$i$ angles of the rapidly rotating stars are colour-coded in units of
degrees. Only single stars are included in the synthetic
populations. The gravity darkening law from \cite{2011A&A...533A..43E}
and the limb-darkening law from \cite{2000A&A...363.1081C} have been
implemented. The synthetic absolute magnitudes are calculated based on
\cite{2018A&A...616A...4E} for \textit{Gaia} Data Release 2. The grey
shaded area in Figure \ref{fig:test_angle} shows the region that
covers the mass range of $\unit[1.8]{M_\odot}<M<\unit[3.6]{M_\odot}$, which is similar to the shaded region in
Figure~\ref{fig:RV_samples} that represents the loci where the split
pattern is evident. The effect of varying $i$ on the morphology of the
rapidly rotating population is significant in the region close to the
MSTO, while it does not extend the full width of the rMS.

Gravity darkening caused by stellar rotation makes the poles hotter
than the equators at the surface of rapidly rotating stars. The
contrast of the effective temperature between the poles and equators
could be enhanced by the limb-darkening effect
\citep{2014A&A...566A..21G}. This results in a scenario where rapidly
rotating stars observed pole-on appear hotter and brighter than their
counterparts which are observed equator-on
\citep{2014A&A...566A..21G}, i.e., they will be bluer and brighter in
the CMDs of star clusters. The direction of this shift seems to align
with the distribution of the rMS (see the rapidly rotating populations
for different $i$ and the shift of a star for different $i$ in Figure
\ref{fig:test_angle}), making rapidly rotating stars with small $i$,
which correspond to small $v\sin i$, appear on the rMS alongside their
counterparts with large $i$. In Figure \ref{fig:test_angle_old_age},
we also plot the distributions of synthetic populations with older
ages ($\unit[150]{Myr}$ and $\unit[200]{Myr}$), where we find a
similar phenomenon, i.e., that rapid rotators with different $i$ are
located along the rMSs. The physics that causes this phenomenon is
unclear. It might be linked to the correlation of effective
temperature with luminosity at the surface towards the observer at
different inclination angles, or to the profiles of the colours and
magnitudes observed through \textit{Gaia} passbands. The rMS stars
with small $v\sin i$ observed in NGC 2422 might thus be some rapidly
rotating single stars showing their poles, which breaks the
$v\sin{i}$--colour correlation expected for the upper MS stars in
young star clusters.

\begin{figure*}
    \includegraphics[width=0.6\textwidth]{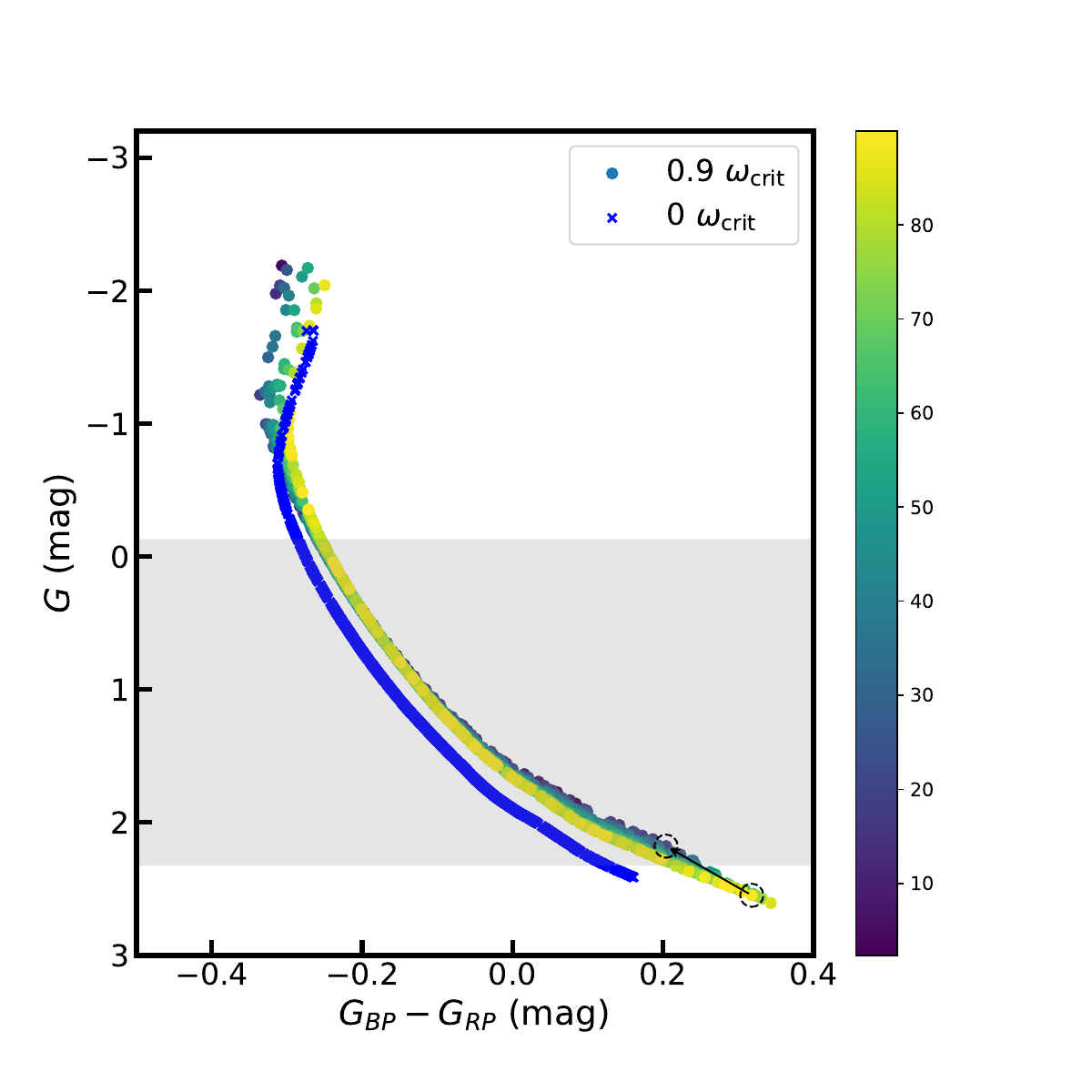}  \\
    \caption{Synthetic coeval populations with different stellar
      rotation rates derived from the SYCLIST database
      \citep{2013A&A...553A..24G, 2014A&A...566A..21G}. The blue
      crosses show the non-rotating population. The dots show the
      rapidly rotating population at $0.9 \mathrm{\omega_{crit}}$,
      colour-coded by the inclination angles of the stellar rotation
      axes, in units of degrees. The dashed circles show the loci of
      two stars with the same mass ($\unit[1.72]{M_{\odot}}$) but
      different inclination angles, $i$. The arrow represents the
      shift in colour and magnitude from $i=\unit[89]{^{\circ}}$ to
      $i=\unit[19]{^{\circ}}$. The grey shading covers the mass range of  $\unit[1.8]{M_\odot}<M<\unit[3.6]{M_\odot}$.}\label{fig:test_angle}
\end{figure*}

\begin{figure*}
\begin{tabular}{cc}
    \includegraphics[width=0.5\textwidth]{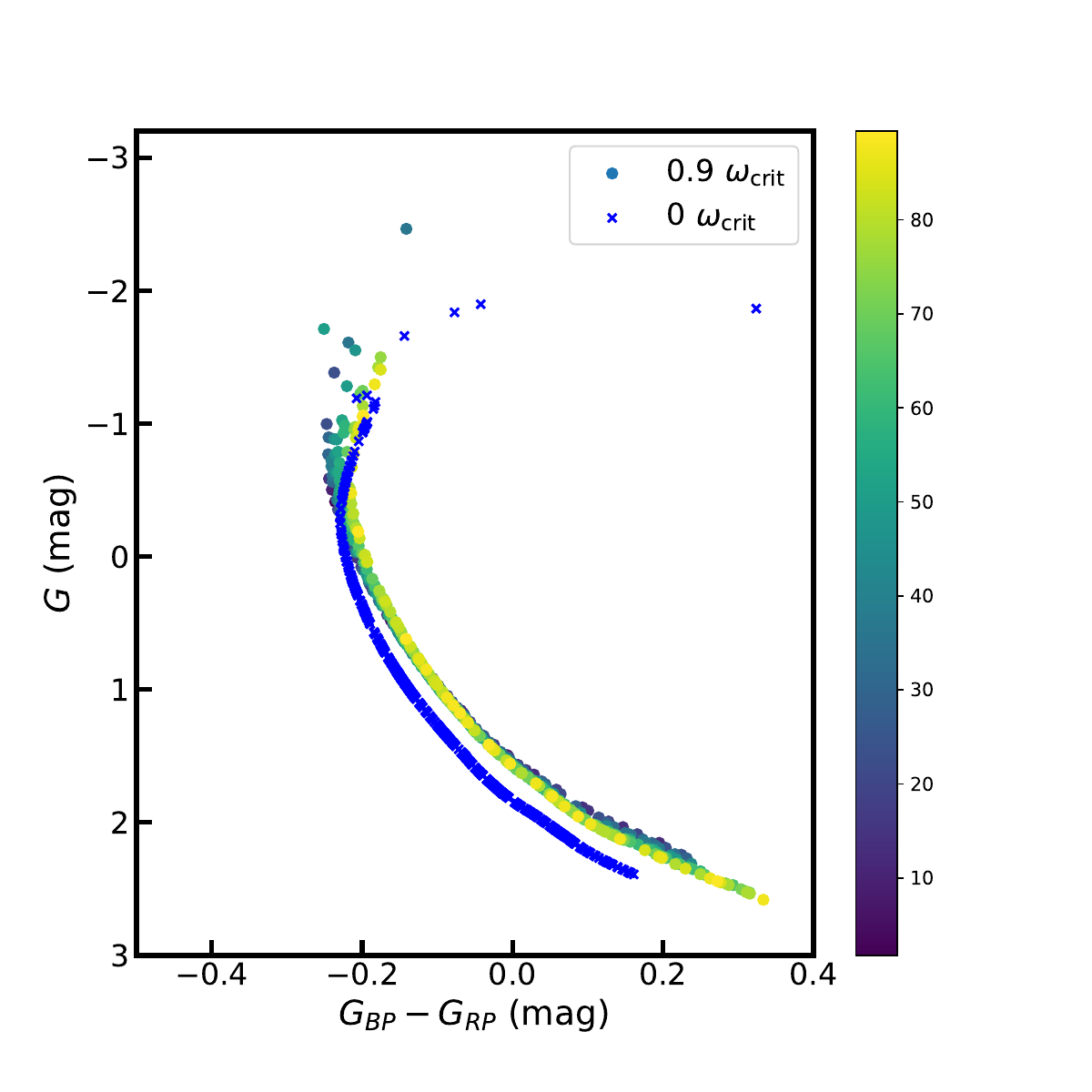}  & \includegraphics[width=0.5\textwidth]{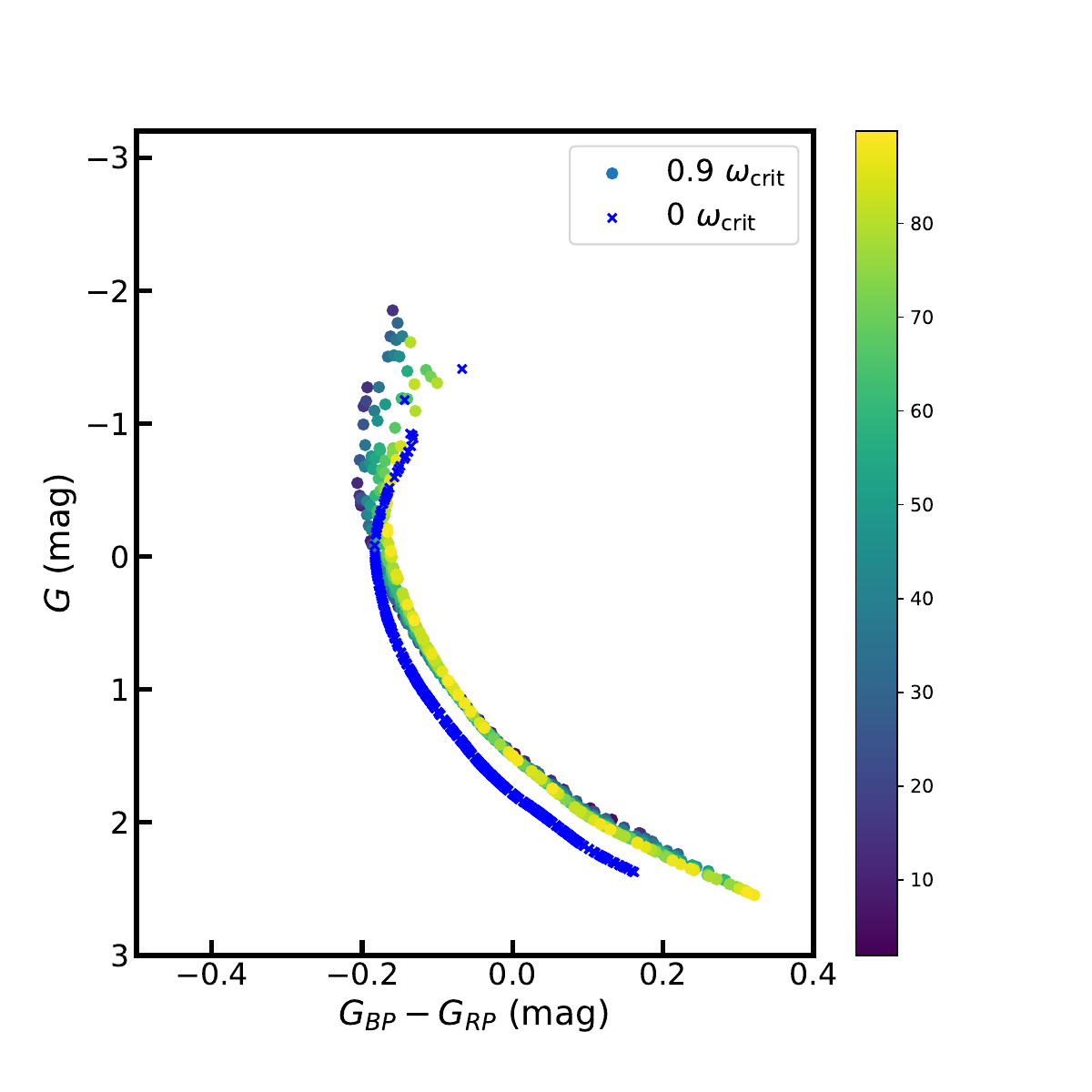} \\
\end{tabular}
	\caption{As Figure \ref{fig:test_angle}, but for synthetic
          populations with ages of $\unit[150]{Myr}$ (left) and
          $\unit[200]{Myr}$ (right).} \label{fig:test_angle_old_age}
\end{figure*}

\section{Conclusion}\label{sec:conclusion}

Using high-resolution spectra observed with the CFHT at multiple
epochs, we explored the RV variability of 21 slowly rotating stars
($v\sin{i}\leq \unit[104]{km\,s^{-1}}$) along the split MS of NGC 2422
($\unit[90]{Myr}$). We found that (1) most our targets exhibit RV
  variations that are much smaller than those expected for binaries
  that are tidally locked within the cluster age (Case 2-Obs and Case
  3-Obs) or within $\unit[10]{Gyr}$ (Case 2-Obs-p and Case 3-Obs-p);
  (2) The dispersion of the measured RVs at a single epoch is much
  smaller than that for synthetic binaries that become tidally locked
  within the cluster age (Case 1-Obs) or within $\unit[10]{Gyr}$ (Case
  1-Obs-p). We thus conclude that most of our targets are not close
  binaries whose stellar rotation rates can be tidally reduced
  substantially on time-scales shorter than the cluster age, that is,
  tidal interactions are not dominant in the formation of slowly
  rotating stars along the split MSs of young star clusters. However,
  we emphasise that this result is based on theoretical predictions for binary separations to partially or fully synchronise the stars in
  \citetalias{2022ApJ...938...42H} using equation (44) of \cite{2002MNRAS.329..897H}, which is based on the dynamical tides theory of \cite{1975A&A....41..329Z,1977A&A....57..383Z}. Observations by
  \cite{2004ApJ...616..562A} show a stronger effect of tidal
  interactions on stellar rotation than the predictions from the above
  theory on shorter times-scales or larger binary separations. The
  latter authors found that the rotation rates of the primary
  B0--F0-type stars in binaries with $T_{\mathrm{orb}}$ of 4--500 days
  are also substantially slowed down compared with those of single
  stars. If our targets are in binaries with
  $\unit[100]{days}<T_{\mathrm{orb}}<\unit[500]{days}$, they may show
  small RV variations like those observed. Based on
  \cite{2004ApJ...616..562A}, the stellar rotation rates in such
  systems can be reduced significantly by tidal interactions. 

The slow rotators along the split MS could be single stars or wide
binaries whose separations are larger than $\sim \unit[0.6]{AU}$
  ($T_{\mathrm{orb}}>\sim \unit[100]{days}$). This kind of binary
systems could show small RV variations like those of most our targets
in time-domain observations. In \citetalias{2022ApJ...938...42H}, we
detected a large fraction of rMS stars showing similar $v\sin i$ to
those of the bMS stars. They might be photometric binaries with
separations larger than $\sim \unit[0.6]{AU}$ and smaller than
$\sim \unit[333]{AU}$. The photometry of rapidly rotating single stars
with small inclination angles could also account for the rMS stars
that exhibit small projected rotation rates.

\section*{acknowledgements}
This research uses data obtained through the Telescope Access Program
(TAP) of China. This work has made use of data from the European Space
Agency (ESA) mission {\it Gaia}
(\url{https://www.cosmos.esa.int/gaia}), processed by the {\it Gaia}
Data Processing and Analysis Consortium (DPAC,
\url{https://www.cosmos.esa.int/web/gaia/dpac/consortium}). Funding
for the DPAC has been provided by national institutions, in particular
the institutions participating in the {\it Gaia} Multilateral
Agreement. This research has used the POLLUX database
(http://pollux.oreme.org), operated at LUPM (Universit\'e
Montpellier--CNRS, France), with the support of the PNPS and
INSU. This work was supported by the National Natural Science
Foundation of China (NSFC) through grant 12233013 and 12073090. TAP
has been funded by the TAP member institutes. R.d.G. was supported in
part by the Australian Research Council Centre of Excellence for All
Sky Astrophysics in 3 Dimensions (ASTRO 3D), through project number
CE170100013. L.C. acknowledges support from the NSFC through grants
12090040 and 12090042. J.Z. acknowledges NSFC grant 12073060, and the
Youth Innovation Promotion Association, Chinese Academy of Sciences. Z.S. acknowledges support from the NSFC through grants 12273091 and U2031139. We are grateful to the anonymous referee for their very useful
  and important suggestions. C.H. thanks Deepak Chahal (Macquarie
University, Australia) for a discussion about the bifurcation of the
rotation periods of solar- and late-type stars.

\section*{Data Availability}

The public data used in our analysis is accessible through the
following links:

 \renewcommand{\labelitemi}{\textendash}
 \begin{itemize}
    \item \textit{Gaia} EDR3 \citep{2016A&A...595A...1G,2021A&A...649A...1G}: \url{https://gea.esac.esa.int/archive/}
    \item PARSEC Isochrones \citep{2017ApJ...835...77M}: \url{http://stev.oapd.inaf.it/cgi-bin/cmd}
    \item SYCLIST Database \citep{2013A&A...553A..24G, 2014A&A...566A..21G}: \url{ https://www.unige.ch/sciences/astro/evolution/en/database/}
\end{itemize}

The spectroscopic data used in this paper can be shared based on
reasonable request to the corresponding author.

\bibliographystyle{mnras}
\bibliography{paper.bib}






\bsp	
\label{lastpage}
\end{document}